\documentclass[conference,letterpaper]{IEEEtran}

\IEEEoverridecommandlockouts  

\usepackage[utf8]{inputenc} 
\usepackage[T1]{fontenc}
\usepackage{url}
\usepackage{ifthen}
\usepackage{cite}
\usepackage[cmex10]{amsmath} 
\usepackage{xcolor}
\usepackage{amssymb}
\usepackage{graphicx}
\usepackage[font=scriptsize]{caption}
\usepackage{xfrac}

\usepackage{graphicx}
\usepackage{hyperref} 
\usepackage{amsmath} 
\usepackage{xfrac}
\usepackage{amssymb}
\usepackage{mathtools}
\usepackage{tikz}
\usepackage{booktabs}
\usepackage{makecell}

\usepackage{amsmath, amssymb, algorithm, algpseudocode}
\usepackage{bm}

\usepackage{subcaption}

\usepackage{amsthm}

\usepackage{multirow}
\usepackage[table]{xcolor}

\theoremstyle{remark}
\newtheorem{remark}{Remark}

\usetikzlibrary{positioning}
\usetikzlibrary{trees}
\usetikzlibrary{shapes,arrows}
\usetikzlibrary{arrows}
\usetikzlibrary{arrows,positioning, calc}

\newcommand{\tabref}[1]{Tab.~\ref{#1}} 
\newcommand{\figref}[1]{Fig.~\ref{#1}}  
\newcommand{\secref}[1]{Sec.~\ref{#1}}  
\renewcommand{\algref}[1]{Alg.~\ref{#1}}  

\begin{document}
\title{Residual Diffusion Models for Variable-Rate Joint Source Channel Coding of MIMO CSI} 

\author{%
  \IEEEauthorblockN{
    Sravan Kumar Ankireddy\IEEEauthorrefmark{2},
    Heasung Kim\IEEEauthorrefmark{2},
    Joonyoung Cho\IEEEauthorrefmark{3},
    Hyeji Kim\IEEEauthorrefmark{2}
  }
  \IEEEauthorblockA{\IEEEauthorrefmark{2} The University of Texas at Austin}
  \IEEEauthorblockA{\IEEEauthorrefmark{3}Samsung Research America}

  \thanks{This paper was presented in part at IEEE Asilomar Conference on Signals, Systems \& Computers, Pacific Grove, CA, Oct 2025.~\cite{ankireddy2025residual}}
  
  \thanks{Source code available at: \href{https://github.com/UTAustin-ITML/rd-jscc}{github.com/UTAustin-ITML/rd-jscc}.}
  
  \thanks{Correspondence to: \{sravan.ankireddy, hyeji.kim\}@utexas.edu.}
}

 

 
  





\maketitle

\let\emptyset\varnothing

\begin{abstract}

Despite significant advancements in deep learning-based CSI compression, some key limitations remain unaddressed. Current approaches predominantly treat CSI compression as a source-coding problem, thereby neglecting transmission errors. 
Conventional separate source and channel coding suffers from the cliff effect, leading to significant deterioration in reconstruction performance under challenging channel conditions.
While existing autoencoder-based compression schemes can be readily extended to support joint source-channel coding, they struggle to capture complex channel distributions and exhibit poor scalability with increasing parameter count. To overcome these inherent limitations of autoencoder-based approaches, we propose Residual-Diffusion Joint Source-Channel Coding (RD-JSCC), a novel framework that integrates a lightweight autoencoder with a residual diffusion module to iteratively refine CSI reconstruction. Our flexible decoding strategy balances computational efficiency and performance by dynamically switching between low-complexity autoencoder decoding and sophisticated diffusion-based refinement based on channel conditions. Comprehensive simulations demonstrate that RD-JSCC significantly outperforms existing autoencoder-based approaches in challenging wireless environments. Furthermore, RD-JSCC offers several practical features, including a low-latency 2-step diffusion during inference, support for multiple compression rates with a single model, robustness to fixed-bit quantization, and adaptability to imperfect channel estimation.

\end{abstract}


\section{Introduction}\label{sec:intro}
As data transmission volumes continue to increase, massive multiple-input multiple-output (MIMO) has emerged as a fundamental technology for scaling next-generation wireless networks. By employing a large array of antennas at the base station (BS), multiple user equipment (UE) can achieve high-throughput communication, even under suboptimal channel conditions. However, achieving this requires accurate channel state information (CSI) to enable effective precoding for downlink transmission. In frequency division duplexing (FDD) systems, the uplink CSI is obtained via channel estimation, while the downlink CSI must be fed back from the user equipment (UE) in an efficient manner~\cite{sim2016compressed}.


Deep learning (DL) has significantly advanced various areas within physical layer communication, including non-linear channel code design~\cite{kim2018communication,makkuva2021ko,jamali2022productae,ankireddy2024nested,ankireddy2025lightcode}, neural channel decoding~\cite{nachmani2016learning,shlezinger2020viterbinet,choukroun2022error,hebbar2022tinyturbo, ankireddy2023interpreting, hebbar2024deeppolar}, and MIMO channel estimation~\cite{wen2018deep, chun2019deep, soltani2019deep}. This work focuses specifically on the challenge of \textit{lossy compression} of CSI at the physical layer. Traditional compression techniques, such as compressed sensing~\cite{kuo2012compressive}, do not work well CSI compression due to the lack of inherent sparsity in CSI structures, making deep learning a better alternative. 
The field of lossy compression using neural networks, commonly referred to as \textit{neural lossy compression}, has gained considerable attention in applications such as image compression~\cite{balle2016end,balle2018variational,li2023task} and video compression~\cite{li2023neural}. More recently, similar methodologies have been leveraged to significantly enhance the efficiency of CSI compression~\cite{guo2022overview}, starting with CSINet~\cite{wen2018deep}, which achieved substantial improvements over the then state-of-the-art compressed sensing methods by leveraging convolutional neural networks (CNNs). This breakthrough led to a series of subsequent studies that further refined CNN-based compression techniques~\cite{wang2019deep,li2020novel,liu2021hyperrnn,lu2020multi, kim2022learning}. Furthermore, similar approaches have been adapted for joint source-channel coding (JSCC) of CSI~\cite{xu2022deep}.


Recently, neural image compression has witnessed a significant breakthrough in both compression efficiency and reconstruction quality with the adoption of diffusion models~\cite{ho2020denoising,song2020denoising}. Originally designed to generate novel images based on various conditioning variables, these models were rapidly adapted for image compression. 
In~\cite{yang2023lossy}, the authors introduced a diffusion-based compression framework that reconstructs images through a reverse diffusion process conditioned on contextual information, outperforming certain GAN-based methods. Furthermore, extremely low-rate compression schemes have been proposed by leveraging the strong image priors of pretrained diffusion models~\cite{careil2023towards}. Likewise,~\cite{mao2024extreme} utilized vector quantized generative adversarial networks (VQGANs), demonstrating the effectiveness of robust codebook priors for discrete visual feature representation in image synthesis, thus providing new insights for compression. More recently,~\cite{li2023task} applied principal component analysis (PCA) to enable variable-rate compression, while~\cite{li2024once} proposed leveraging spatial entropy~\cite{celik2014spatial} to learn latent representations at multiple resolutions, facilitating a continuous range of bitrates and~\cite{song2024variable} proposed a dictionary-based quantization approach with variable length masks to support multiple compression rates.



While diffusion models have been studied extensively in image compression, their potential for CSI compression remains underexplored. A notable advancement in this direction is the generative diffusion-based CSI compression framework proposed in~\cite{kim2025generative}, demonstrating significant improvements over traditional autoencoder-based methods for modeling complex channels. Building upon this promising foundation, our work aims to further enhance diffusion-based CSI compression in two key dimensions. First, we prioritize reconstruction fidelity by balancing output diversity, shifting the model's objective from generating novel samples towards maximizing reconstruction accuracy. Second, we employ a nested representation learning approach to capture multiple compression levels within a unified latent space. By using continuous-valued latents with scalar quantization, our method naturally handles noisy transmission, as it maps directly to subcarriers with power constraints. This design enables dynamic and graceful adaptation of the compression rate in response to fluctuating bandwidth constraints.

In this work, we propose a novel residual diffusion-based variable-rate CSI compression framework tailored for efficient compression of CSI matrices in massive MIMO systems. The proposed framework features an encoder leveraging a low-complexity convolutional neural network (CNN) architecture and a two-stage decoding process at the receiver. Specifically, the decoder comprises an initial CNN-based reconstruction stage followed by a U-Net-based diffusion refinement model, which iteratively enhances CSI reconstruction quality. The main contributions of this paper are summarized as follows:

\begingroup

\begin{itemize}
    

    \item We propose RD-JSCC, a flexible two-stage framework that initializes reverse diffusion with a coarse autoencoder estimate and iteratively refines the residual error. This design enables adaptive switching between lightweight autoencoder decoding and 2-step diffusion-based refinement to balance reconstruction quality with low-latency inference requirements.(\secref{sec:resdiff})
    
    \item We adopt a nested representation-learning framework based on continuous-valued latent spaces that supports multiple compression rates within a single unified model, enabling fine-grained bandwidth adaptation without retraining or model-switching overhead (\secref{sec:var_rate}).
    
    \item We validate RD-JSCC against state-of-the-art deep learning-based JSCC schemes using the COST2100 outdoor and 3GPP indoor datasets, demonstrating order-of-magnitude performance improvements over non-diffusion baselines and establishing that diffusion-based refinement is particularly effective for statistically complex channel distributions (\secref{sec:results}).
    
    \item We conduct comprehensive ablation studies that quantify the contributions of the residual diffusion formulation, encoder complexity, and SNR adaptation, and systematically analyze the complexity-performance trade-offs across different channel scenarios to determine when diffusion-based refinement is justified (\secref{sec:model_choice}).
    
    \item We address practical deployment considerations by demonstrating that NMSE improvements translate to substantial BLER reductions, validating robustness to imperfect channel estimation, and demonstrating fixed-bit quantization down to 4 bits per latent dimension with negligible performance degradation (\secref{sec:bler}, \secref{sec:che}, \secref{sec:quant}).
\end{itemize}
\endgroup

\section{System Model and Problem Formulation}\label{sec:sysmodel}

In this work, we consider a massive MIMO system operating in frequency-division duplex (FDD) mode, in which a base station (BS) with $N_t$ antennas communicates with user equipment (UE) with $N_r$ antennas. Following conventions in existing literature and available datasets, we consider massive MIMO systems where $N_t \gg 1$ and set $N_r = 1$, resulting in a SIMO configuration from each user's perspective. This single-antenna receiver assumption simplifies CSI feedback while maintaining practical relevance, as the base station can still serve multiple users simultaneously through spatial multiplexing. Our proposed techniques readily extend to full MIMO scenarios with $N_r > 1$ by incorporating an additional antenna dimension in the deep learning architecture. 
The downlink CSI is $\bm{H}_d \in \mathbb{C}^{N_c \times N_t}$ and the uplink CSI is denoted by $\bm{H}_u \in \mathbb{C}^{N_c \times N_t}$ in the spatial-frequency domain, where $\mathbb{C}$ denotes the set of complex numbers and $N_c$ is the number of subcarriers.

The encoder $f_{\text{enc}}$ at the UE is designed to efficiently compress the high-dimensional channel measurement $\bm{H}_d$ into a fixed-length representation $\mathbf{s} \in \mathbb{C}^k$. The compression rate is thus given by
\begin{equation*}
    \rho = \frac{k}{N_t \times N_c}.
\end{equation*}
The compressed representation can be transmitted on a noisy uplink channel using $k$ subcarriers, while imposing a unit power constraint for the subcarriers $\frac{1}{k}\,\mathbb{E}\bigl[\mathbf{s}\,\mathbf{s}^*\bigr] = 1$. The received signal at the BS is processed using maximal ratio combining (MRC).

The compressed representation at the receiver is processed in two stages. First, the decoder $f_{\text{dec}}$ reconstructs an estimate $\hat{\bm{H}}_d$ from the compressed representation $\mathbf{s}$, aiming to minimize distortion relative to the original input $\bm{H}_d$. Next, to further refine the reconstruction, the noisy estimate $\hat{\bm{H}}_d$ is processed by a denoising diffusion model $f_{\text{den}}$, iteratively reducing the distortion using the reverse diffusion. Specifically, we use the residual diffusion formulation~\cite{liu2024residual}.
Following standard practice in CSI compression literature, we adopt mean squared error (MSE) as the distortion metric.

The encoder function is defined as $f_{\text{enc}}: \bm{H}_d \mapsto \mathbf{s}$, while the decoder function is given by $f_{\text{dec}}: \mathbf{s} \mapsto \hat{\bm{H}}_d$, and the denoising model operates as 
$f_{\text{den}}: \hat{\bm{H}}_d \mapsto \bm{H}'_d$.
Given the encoder parameters $\theta_{\text{enc}}$, the compressed representation $\mathbf{s}$ is obtained as $\mathbf{s} = f_{\text{enc}}(\bm{H}_d; \theta_{\text{enc}})$. Similarly, the decoder, parameterized by $\theta_{\text{dec}}$, reconstructs an estimate of the target as $\hat{\bm{H}}_d = f_{\text{dec}}(\mathbf{s}; \theta_{\text{dec}})$. Finally, the denoising model, governed by $\theta_{\text{den}}$, further enhances the reconstruction, producing $H'_d = f_{\text{den}}(\hat{\bm{H}}_d; \theta_{\text{den}})$. The complete set of model parameters is thus given by $\theta = (\theta_{\text{enc}}, \theta_{\text{dec}}, \theta_{\text{den}})$.

The learning process is designed to minimize the following objective function:
\begin{equation}
\begin{split}
\theta^* = \arg\min_{\theta} \mathbb{E}_{p(\bm{H}_d, \bm{H}'_d)} \big[ & d\big(\bm{H}_d, f_{\text{den}}(f_{\text{dec}}(f_{\text{enc}}(\bm{H}_d; \theta_{\text{enc}}); \\
& \theta_{\text{dec}}); \theta_{\text{den}})\big) \big]
\end{split}
\end{equation}

where $d(\cdot, \cdot)$ represents a distortion metric that quantifies reconstruction quality. The parameters $\theta = (\theta_{\text{enc}}, \theta_{\text{dec}}, \theta_{\text{den}})$ are optimized using a two-stage training process, detailed in~\secref{sec:resdiff}.

\section{Deep Learning for CSI compression}\label{sec:deep_csi}

In this section, we introduce a low-complexity autoencoder-based solution for CSI compression.
By considering the task of CSI compression as an image compression problem, several works based on an autoencoder structure were proposed~\cite{wen2018deep,lu2018mimo,guo2020convolutional,lu2020multi,hu2021mrfnet,chen2021high,cao2021lightweight,ji2021clnet}. Notable works among them include CSINet~\cite{wen2018deep}, which was one of the first deep learning approaches proposed and outperformed compressed sensing baselines such as LASSO and BM3D-AMP. CRNet~\cite{lu2020multi} proposed a multi-resolution deep learning framework for CSI feedback in massive MIMO systems, enabling scalable compression across different feedback overheads. Recently, a transformer-based architecture that utilizes stripe-wise spatial features was introduced in~\cite {hu2023csi} to enhance the efficiency of CSI compression in massive MIMO systems.

\begin{figure}[!htb]
    \centering
    \begin{subfigure}[b]{0.95\linewidth}
        \centering
        \includegraphics[width=\linewidth]{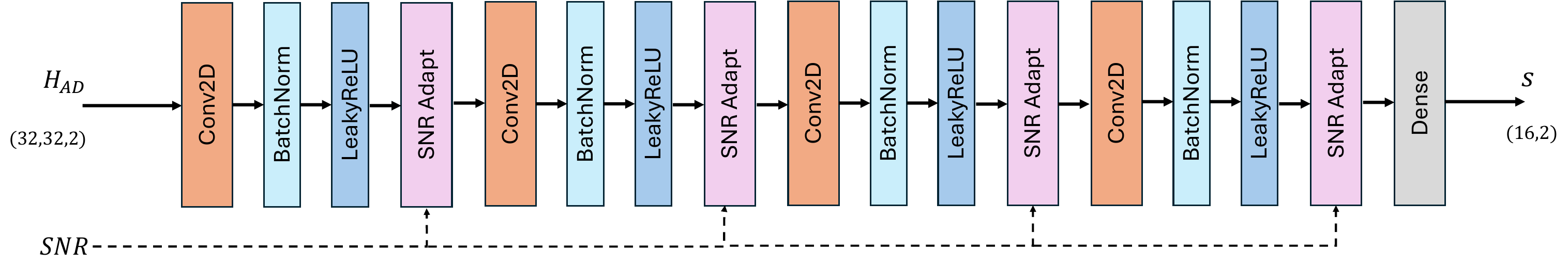}
        \captionsetup{font=small}
        \caption{CNN encoder}
        \label{fig:neural_mod}
    \end{subfigure}
    
    \vspace{1em} 

    \begin{subfigure}[b]{0.95\linewidth}
        \centering
        \includegraphics[width=\linewidth]{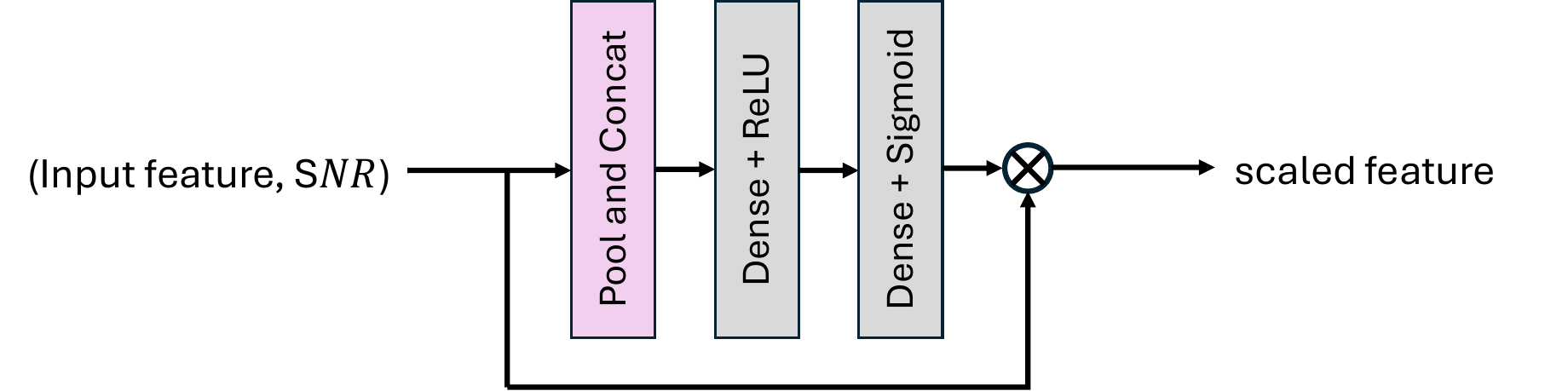}
        \captionsetup{font=small}
        \caption{Feature scaling for SNR adaptation}
        \label{fig:SNR_Adapt}
    \end{subfigure}
    
    \captionsetup{font=small}
    \caption{Low-complexity CSI encoder.}
    \label{fig:encoder}
\end{figure}

Given the three-dimensional, spatially correlated structure of the CSI matrix, a convolution-based architecture is an efficient choice for input compression. 
We chose a low-complexity design for the auto-encoder. The encoder is implemented as lightweight CNN layers followed by a fully connected layer. To maintain robustness under varying signal-to-noise ratio (SNR) conditions, we employ an SNR-adaptation module that dynamically scales feature activations based on SNR. The complete architecture of both the encoder and the SNR-adaptation block is shown in~\figref{fig:encoder}.

We employ continuous latent representations instead of vector quantization (VQ) because VQ suffers from the cliff effect: single-bit errors in codebook indices can cause catastrophic reconstruction failure when transmitted over noisy channels. Continuous representations provide graceful degradation: as channel noise increases, reconstruction errors scale proportionally, thereby supporting joint source-channel coding without requiring additional error protection layers. This approach enables direct modulation onto uplink subcarriers while maintaining compatibility with scalar quantization when needed, as demonstrated in Section~\ref{sec:quant}.

\begin{figure}[!htb]
    \centering
    \begin{subfigure}[b]{0.95\linewidth}
        \centering
        \includegraphics[width=\linewidth]{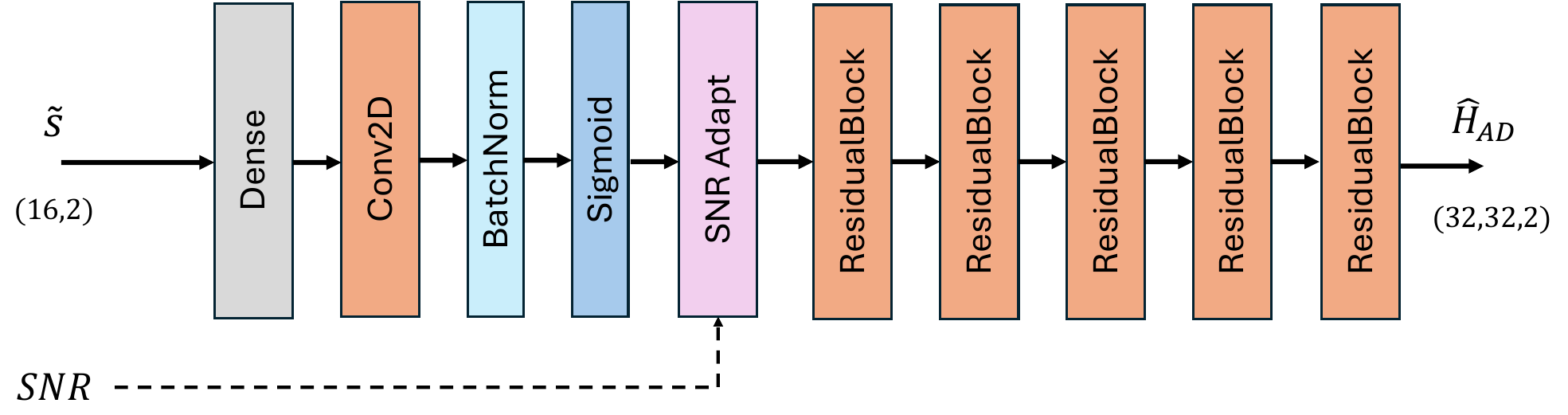}
        \captionsetup{font=small}
        \caption{CNN decoder}
        \label{fig:cnn_dec}
    \end{subfigure}
    
    \vspace{1em} 

    \begin{subfigure}[b]{0.95\linewidth}
        \centering
        \includegraphics[width=\linewidth]{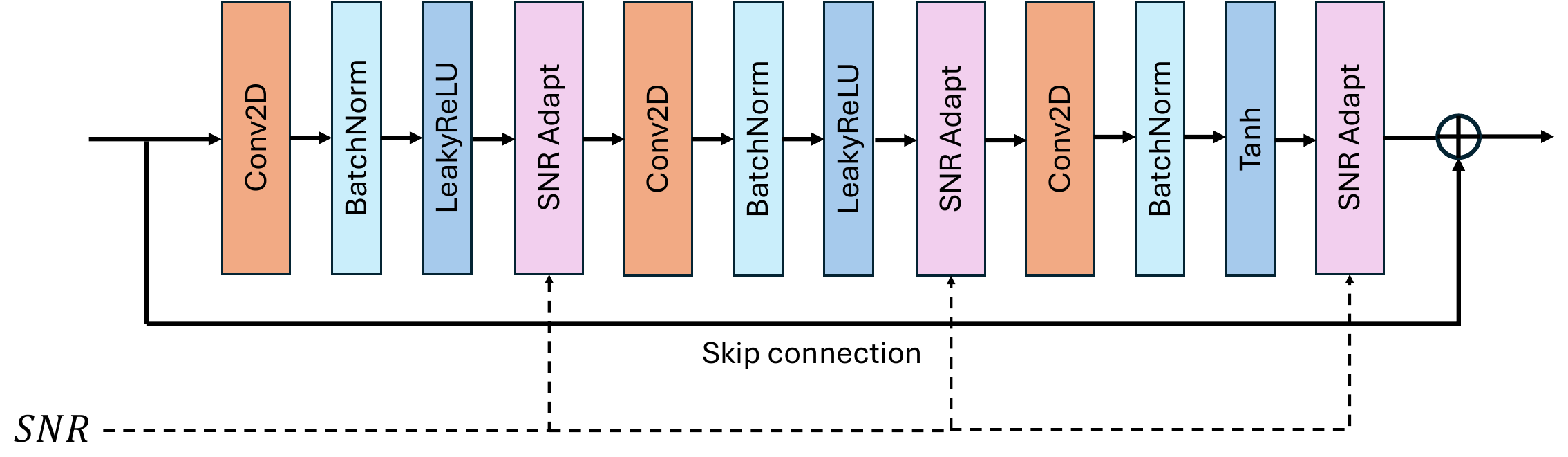}
        \captionsetup{font=small}
        \caption{Residual connection}
        \label{fig:res_block}
    \end{subfigure}
    
    \captionsetup{font=small}
    \caption{Low-complexity CSI decoder.}
    \label{fig:decoder}
\end{figure}

The decoder employs a moderately deeper architecture to extract richer representations. It starts with a fully connected layer, followed by an initial convolutional layer and a stack of five residual blocks that enhance feature propagation and stabilize training. An identical SNR-adaptation module is employed in the decoder, mirroring the encoder design. A schematic of the decoder, including the residual block layout, is provided in~\figref{fig:decoder}.  

We train the autoencoder in a supervised manner to learn an effective low-complexity compression scheme. The encoder compresses the continuous channel measurement $\bm{H}_d$ into a compact latent vector $\mathbf{s} \in \mathbb{C}^k$. The decoder reconstructs the approximate channel measurement $\hat{\bm{H}}_d$. The encoder and decoder are jointly optimized end-to-end to minimize the CSI reconstruction loss. We use the MSE loss to enforce reconstruction fidelity by ensuring that the reconstructed output closely matches the input, given by:
\begin{equation}
    \mathcal{L}_{\text{MSE}}(\bm{H}_d, \hat{\bm{H}}_d) = \|\bm{H}_d - \hat{\bm{H}}_d\|^2, 
    \label{eq:loss_mse}
\end{equation}
where $\|\cdot\|$ denotes the Frobenius norm.

Several works have demonstrated that CNN–based autoencoders can effectively minimize the reconstruction NMSE for both standalone CSI compression \cite{wen2018deep,hu2023csi} and joint source–channel coding of CSI \cite{xu2022deep}. Although this supervised formulation suffices for relatively simple channel distributions, it struggles when the underlying channel distribution becomes more complex. Recent work on CSI compression over the COST2100 outdoor channel \cite{kim2025generative} demonstrated that generative diffusion models significantly outperform conventional CNN autoencoders in such challenging settings. Motivated by these findings, we augment our autoencoder with a diffusion refinement stage at the base station. This hybrid design invokes the diffusion module only when the channel complexity warrants it, thereby delivering superior reconstruction quality without incurring unnecessary computational overhead in simpler channel conditions.


\section{Residual Diffusion for CSI Enhancement}\label{sec:resdiff}

\begin{figure*}[!htb]
    \centering
 	\includegraphics[width=0.95\linewidth]{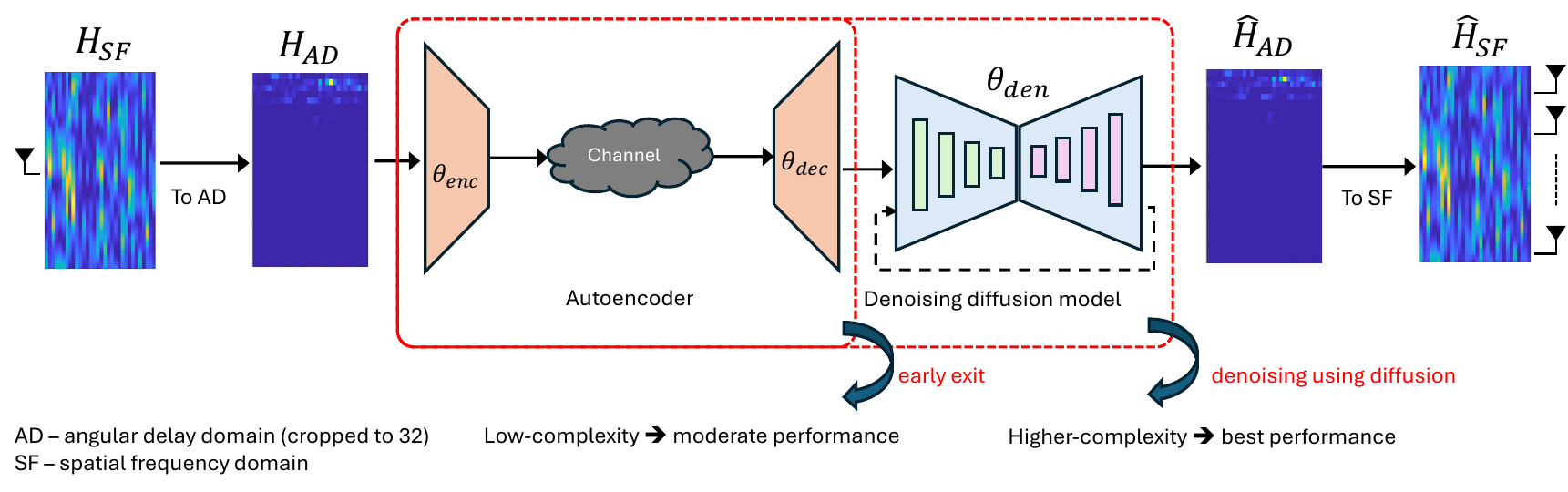}
 	\captionsetup{font=small}
 	\caption{A lightweight autoencoder compresses the MIMO CSI matrix into a low-dimensional latent representation, which is transmitted over a noisy channel. Decoding occurs in two stages: (1) a low-complexity decoder produces a coarse reconstruction, and (2) a diffusion-based denoising model refines the output for enhanced quality.}
 	\label{fig:sysmodel}
\end{figure*}

In this section, we introduce a framework for enhancing CSI reconstruction at the receiver using residual diffusion~\cite{liu2024residual}. Unlike conventional generative diffusion-based approaches, the proposed method can be seamlessly integrated with both learning-based and non-learning-based techniques, as the diffusion module is trained as a standalone denoising model, thereby facilitating practical adaptation. 

\textit{Preprocessing.\ } To reduce computational complexity, we first transform the complex matrix $\bm{H}_d$ from the spatial-frequency domain to the angular-delay domain by applying a two-dimensional inverse fast Fourier transform (2D IFFT). This transformation exploits the inherent sparsity in the angular-delay domain, supported by established assumptions~\cite{wang2018spatial}. Subsequently, we retain only the first $32$ elements along the delay dimension, as the remaining coefficients typically approach zero, yielding compact angular-delay domain representations of $\bm{H}_d$. The original CSI matrices can be reconstructed by appending zero matrices of size $32 \times (N_c-32)$ and performing a 2D FFT. This preprocessing procedure is widely recognized as an efficient technique for CSI representation~\cite {wen2018deep, wang2018deep, lu2020multi}. The performance evaluations reported in this work are conducted in terms of normalized mean squared error (NMSE), following the standard practice in CSI compression literature. 

The receiver first obtains a coarse estimate of the channel using the autoencoder described in~\secref{sec:deep_csi}, which we refer to as Stage 1. Unlike CSI compression methods that use vector quantization to discretize the latent representation into fixed-bit codewords, we map the channel directly to a fixed-length, continuous-valued latent vector. This continuous representation enables direct mapping of each latent dimension to uplink Orthogonal Frequency-Division Multiplexing (OFDM) subcarriers for feedback transmission from the UE to the base station, and naturally captures complex channel impairments, such as multipath fading, within the feedback channel.

In Stage~2, the denoising process leverages the autoencoder output to initialize the reverse diffusion procedure with the coarse CSI estimate. We adopt the U-Net architecture~\cite{ronneberger2015u} as the backbone of the denoising network, following standard practice in generative modeling. Unlike recent generative diffusion-based CSI reconstruction methods~\cite{kim2025generative} that initialize from random Gaussian noise, we employ a residual diffusion approach. This method initializes the reverse diffusion with the autoencoder's coarse estimate and iteratively refines the CSI, as illustrated in Figure~\ref{fig:sysmodel}.
The decoder adaptively switches between 1-stage and 2-stage decoding based on the uplink channel quality. When the uplink SNR exceeds a predetermined threshold, indicating favorable channel conditions, the decoder terminates at Stage~1. Otherwise, Stage~2 diffusion refinement is invoked to further enhance reconstruction quality.


While initializing with random noise facilitates the generation of diverse samples from the underlying distribution, an objective well-suited for data synthesis, it is suboptimal for compression and reconstruction tasks, where compressed latent features already provide a strong prior. Residual diffusion addresses this by starting from a coarse channel estimate, leading to more accurate reconstructions. The effectiveness of residual diffusion for reconstruction tasks has been well demonstrated in the context of image compression~\cite{li2024diffusion}. We now formally describe the complete CSI compression and reconstruction pipeline.

In Stage~1, the encoder produces a compressed representation $\mathbf{s}$, which is then used to obtain a coarse reconstruction $\hat{\bm{H}}_d$ of the input channel $\bm{H}_d$. During the denoising stage, the objective is to further refine $\hat{\bm{H}}_d$ by sampling 
$\mathbf{Z} \sim p(\mathbf{z} \mid \hat{\bm{H}}_d)$. This leads to the formulation of a residual denoising diffusion process, given by
\begin{equation}
    p(\mathbf{z}_{0:T} \mid \hat{\bm{h}}_d) = p(\mathbf{z}_T) \prod_{t=1}^{T} \mathcal{N}(\mathbf{z}_{t-1}; \mu_{\theta}(\mathbf{z}_t, \hat{\bm{h}}_d, t), \beta_t \mathbf{I}),
\end{equation}
where $\hat{\bm{h}}_d$ is the current sample of the coarse estimate of the channel. $\mathbf{z}_{0:T} = (\mathbf{z}_0, \dots, \mathbf{z}_T)$ denotes a realization of the stochastic process $(\mathbf{Z}_0, \dots, \mathbf{Z}_T)$, $\mu_{\theta}$ is the learnable mean function parameterized by $\theta$, $\beta_t$ defines the variance schedule at each time step $t$, and $\mathbf{I}$ is the identity matrix.

In a generative diffusion-based approach, the forward process is defined by progressively adding Gaussian noise with variance $\beta_t \in (0,1)$ to the clean latent features $\mathbf{z}_0$ according to a predefined schedule, as given by
\begin{equation}
    \mathbf{z}_t = \sqrt{\bar{\alpha}_t} \mathbf{z}_0 + \sqrt{1 - \bar{\alpha}_t} \boldsymbol{\epsilon}_t, \quad t = 1,2,\dots,T,
\end{equation}
where $\boldsymbol{\epsilon}_t \sim \mathcal{N}(0, I)$, $\alpha_t = 1 - \beta_t$, and $\bar{\alpha}_t = \prod_{i=1}^{t} \alpha_i$. As $t$ increases, the noisy latent variable $\mathbf{z}_t$ gradually converges to a standard Gaussian distribution. Typically, $T$ is chosen as a relatively large value (e.g., 20–50), with the reverse diffusion process initialized from pure Gaussian noise. However, in the context of CSI compression, this strategy is suboptimal, as a coarse channel estimate $\hat{\bm{H}}_d$ can be readily obtained using either conventional compression techniques~\cite{sim2016compressed} or low-complexity deep learning-based methods~\cite{sun2024efficient}.

To exploit the availability of the coarse estimate $\hat{\bm{H}}_d$, we adopt a residual diffusion approach as presented in~\cite{li2024diffusion}, which modifies the initialization step as
\begin{equation}
    \mathbf{z}_N = \sqrt{\bar{\alpha}_N} \hat{\bm{h}}_d + \sqrt{1 - \bar{\alpha}_N} \boldsymbol{\epsilon}_N,
\end{equation}
where \(N \ll T\). This leads to the following residual diffusion formulation:
\begin{equation}
    \mathbf{z}_t = \sqrt{\bar{\alpha}_t} \left( \mathbf{z}_0 + \eta_t \mathbf{r} \right) + \sqrt{1 - \bar{\alpha}_t} \boldsymbol{\epsilon}_t, \quad t = 1,2,\dots,N,
    \label{eq:z_t}
\end{equation}
where $\mathbf{r}$ denotes the residual between the clean channel $\mathbf{z}_0 = \bm{h}_d$ and the coarse estimate $\hat{\bm{h}}_d$, i.e., $\mathbf{r} = \hat{\bm{h}}_d - \mathbf{z}_0$. The weighting sequence $\{ \eta_t \}_{t=1}^{N}$ is designed such that $\eta_1 \to 0$ and $\eta_N = 1$.

Since the residual $\mathbf{r}$ is not available during inference, residual diffusion assumes a linear relationship among $\mathbf{z}_{t-1}$, $\mathbf{z}_t$, and $\mathbf{z}_0$, analogous to the DDIM framework~\cite{song2020denoising}, given by
\begin{equation}
    \mathbf{z}_{t-1} = k_t \mathbf{z}_0 + m_t \mathbf{z}_t + \sigma_t \boldsymbol{\epsilon},
    \label{eq:z_t_1}
\end{equation}
where $\sigma_t = 0$ for simplicity and $k_t$ and $m_t$ are weighing coefficients from~\cite{li2024diffusion} .Combining~\eqref{eq:z_t} and~\eqref{eq:z_t_1} yields
\begin{equation}
    \frac{\eta_t}{\eta_{t-1}} = \frac{\sqrt{1 - \bar{\alpha}_t} / \sqrt{\alpha_t}}{\sqrt{1 - \bar{\alpha}_{t-1}} / \sqrt{\alpha_{t-1}}}
    \quad \Rightarrow \quad \eta_t = \lambda \frac{\sqrt{1 - \bar{\alpha}_t}}{\sqrt{\alpha_t}},
    \label{eq:lambda}
\end{equation}
where the scaling factor $\lambda$ is set to $\frac{\sqrt{\bar{\alpha}_N}}{\sqrt{1 - \bar{\alpha}_N}}$ to satisfy the condition $\eta_N = 1$.

Substituting~\eqref{eq:lambda} into~\eqref{eq:z_t}, the forward diffusion process can be expressed as
\begin{equation}
    \mathbf{z}_t = \sqrt{\bar{\alpha}_t} \left( \mathbf{z}_0 + \lambda \frac{\sqrt{1 - \bar{\alpha}_t}}{\sqrt{\bar{\alpha}_t}} \mathbf{r} \right) + \sqrt{1 - \bar{\alpha}_t} \boldsymbol{\epsilon}_t,
\end{equation}
which simplifies to
\begin{equation}
    \mathbf{z}_t = \sqrt{\bar{\alpha}_t} \mathbf{z}_0 + \sqrt{1 - \bar{\alpha}_t} (\lambda \mathbf{r} + \boldsymbol{\epsilon}_t).
\end{equation}
Ultimately, the denoising network is trained to recover the clean channel $\mathbf{z}_0$ from noisy observations at various noise levels encountered during the forward diffusion process. The corresponding diffusion loss is formulated as
\begin{equation}
    \mathcal{L}_{\text{diff}} = \mathbb{E}_{\mathbf{Z}_0, \mathbf{T}} \left[ \frac{\bar{\alpha}_\mathbf{T}}{1 - \bar{\alpha}_\mathbf{T}} \left\| \mathbf{Z}_0 - f_{\text{den}}(\mathbf{Z}_\mathbf{T}, \hat{\bm{H}}_d, \mathbf{T}; \theta_{\text{den}}) \right\|^2 \right],
    \label{eq:loss_diff}
\end{equation}
where $f_{\text{den}}(\cdot)$ is the denoising network parameterized by $\theta_{\text{den}}$.

\vspace{5 pt}
\textbf{Diffusion with $\chi$-prediction.\ } In Equation~\eqref{eq:loss_diff}, the loss is computed between the clean image and the model's prediction, compelling the network to directly predict the image rather than the conventional approach of predicting the added noise. This technique, known as \textit{$\chi$-prediction}, proves particularly advantageous when operating with a limited number of denoising steps (e.g., $T < 50$). As our experimental results in subsequent sections demonstrate, this approach facilitates low-latency inference by enabling a transition to one-step decoding during inference with minimal or no degradation in reconstruction quality.


\emph{Model Training.\ } We propose a two-stage training strategy to optimize performance while maintaining computational efficiency. In the first stage, the autoencoder model is trained using the MSE loss~\eqref{eq:loss_mse} between the input CSI matrix and the estimated CSI matrix. Due to the relatively small number of parameters and faster convergence, the computational overhead for Stage~1 is minimal. The detailed training procedure for this stage is outlined in~\algref{alg:train_ae}.

\begin{algorithm}
\caption{Training the autoencoder model}
\textbf{Input:} Initial model $(\theta_\text{enc}, \theta_\text{dec})$, Adam optimizer\\
\textbf{Output:} Updated model $(\theta_\text{enc}, \theta_\text{dec})$

\begin{algorithmic}[1]
\For{$i = 0$ to $N_{\text{train}}$}
    \State Sample $\bm{h}_d$ from the channel distribution
    \State $\mathbf{s} = f_{\text{enc}}(\bm{h}_d; \theta_\text{enc})$
    
    \State $\hat{\bm{h}}_d = f_{\text{dec}}(\mathbf{s}; \theta_\text{dec})$
    \State $\mathcal{L}_{\text{MSE}} = \|\bm{h}_d - \hat{\bm{h}}_d\|^2$
    \State Adam$(\theta_\text{enc}, \theta_\text{dec}, \mathcal{L}_{\text{MSE}})$
\EndFor

\end{algorithmic}
\label{alg:train_ae}
\end{algorithm}

In the second stage, the weights of the autoencoder model are frozen, while a residual conditional denoising diffusion model, based on the U-Net architecture, is trained. This training process optimizes the diffusion loss defined in~\eqref{eq:loss_diff}, essentially training a denoising network. The detailed training procedure for this stage is outlined in~\algref{alg:denoising}.  

\begin{algorithm}
\caption{Training the denoising U-Net}
\textbf{Input:} Pretrained $(\theta_\text{enc}, \theta_\text{dec})$. Initial parameters for denoising network $\theta_\text{den}$, variance schedule $\{\bar{\alpha}_t\}_{t=0}^{T}$, Adam optimizer\\
\textbf{Output:} Updated denoising network $\theta_\text{den}$ 

\begin{algorithmic}[1]
\For{$i = 0$ to $N_{\text{train}}$}

    \State Sample $\bm{h}_d$ form the channel distribution
    
    \State $\mathbf{s} = f_{\text{enc}}(\bm{h}_d; \theta_\text{enc})$

    \State $\hat{\bm{h}}_d = f_{\text{dec}}(\mathbf{s}; \theta_\text{dec}); ~~\mathbf{r} = \hat{\bm{h}}_d - \mathbf{z}_0$
    \State  
        $\mathcal{L}_{\text{diff}} = \frac{\bar{\alpha}_t}{1-\bar{\alpha}_t} \big\| \mathbf{z}_0 - f_\text{den}(\sqrt{\bar{\alpha}_t} \mathbf{z}_0 + \sqrt{1 - \bar{\alpha}_t} (\lambda \mathbf{r} + \epsilon_t)) \big\|^2 $

    \State Adam$(\theta_\text{den}, \mathcal{L}_{\text{diff}})$
\EndFor
\end{algorithmic}
\label{alg:denoising}
\end{algorithm}

\section{Variable Rate Compression}\label{sec:var_rate}


In practical wireless communication systems, the available uplink bandwidth for CSI feedback varies dynamically due to factors such as performance requirements and network congestion. Traditionally, deep learning-based compression methods address multiple compression rates by training separate models for each rate. However, this approach imposes significant overhead on resource-constrained user equipment (UE). To mitigate this, recent research has focused on developing single-model solutions that can handle multiple compression rates.

For instance, CSI-PPPNet~\cite{chen2023csi} introduces a one-sided architecture that supports arbitrary compression ratios using a single deep learning model. This design significantly reduces the number of parameters required at the UE, enhancing efficiency and simplifying deployment. Similarly, the Variable-Rate Code (VRC) approach dynamically allocates feedback bits based on estimated distortion, optimizing the trade-off between compression rate and reconstruction quality~\cite{kim2022learning}. 
Additionally, the Variable Code Size Autoencoder (VCSA) framework adjusts codeword lengths during training, enabling flexible compression without retraining for different rates~\cite{song2024variable} but requires perfect transmission of the codeword indices, which is a disadvantage of vector quantization. Instead of a codebook-based approach, continuous-valued latents with scalar quantization naturally handle noisy transmission by mapping directly to subcarriers with power constraints.



\emph{Matryoshka Representation Learning.} Introduced in~\cite{kusupati2022matryoshka}, Matryoshka Representation Learning(MRL) is a representation learning framework designed to support multiple compression rates within a single model. The key idea behind MRL is to learn a single hierarchical representation that can be truncated to different lengths, yielding compressed representations of varying fidelity. This is achieved by partitioning the learned representation vector into nested subsets, such that each prefix subset corresponds to a progressively higher compression rate. Importantly, MRL enforces joint optimization of all nested representations during training, enabling the model to perform well across a broad spectrum of rate-distortion trade-offs without the need for retraining or maintaining multiple models. This design allows for efficient and adaptive deployment in scenarios with dynamic bandwidth or storage constraints, while maintaining representational consistency and strong generalization across different rates.

To enable CSI feedback compression at multiple target bandwidths $\{k_1, k_2, \dots, k_m\}$, the standard reconstruction loss~\eqref{eq:loss_mse} is extended to account for the cumulative reconstruction errors at each compression level. Specifically, the loss function is modified as
\begin{equation}
    \mathcal{L}_{\text{MRL}} = \sum_{i=1}^m \lambda_i \| \bm{H}_d - \hat{\bm{H}}_d^{(k_i)} \|^2,
    \label{eq:loss_mrl}
\end{equation}
where $\hat{\bm{H}}_d^{(k_i)}$ denotes the reconstructed CSI from the compressed representation of dimension $k_i$, and $\lambda_i$ are weighting factors that balance the importance across different rates. We choose $\lambda_i = \sfrac{1}{i}$ by default, assigning lower weights to loss from higher-compression rates.

In~\figref{fig:mrl}, we present results for the model optimized across four feedback bandwidths $k \in \{8, 16, 24, 32\}$, corresponding to compression rates $\rho \in \{\sfrac{1}{128}, \sfrac{1}{64}, \sfrac{1}{42}, \sfrac{1}{32}\}$. The encoder architecture compresses CSI to a maximum complex-valued latent dimension of $32$, with the MRL training objective (Eq.~12) enabling nested prefix truncation to generate latents of dimensions $24$, $16$, and $8$. 
Crucially, while we explicitly optimize for four discrete compression levels, the nested structure inherently supports \emph{all} intermediate rates, specifically, all integer bandwidths $k \in \{8, 9, \ldots, 32\}$, yielding $25$ distinct compression rates from a single unified model. During inference, the encoder adaptively selects the latent dimension based on available bandwidth, achieving an efficient rate-distortion trade-off without model switching overhead.

\section{Experimental Setup and Results}\label{sec:results}

\subsection{Baselines and comparison.}\label{sec:baselines}
To evaluate our diffusion-based approach, we benchmark against ADJSCC~\cite{xu2022deep}, a recent state-of-the-art non-linear transform method. ADJSCC employs a neural network to transform CSI data from the spatial-frequency domain into a low-dimensional representation, thereby bypassing the traditional IFFT approach, before further compression via a secondary network. We also include the deep JSCC variant of CSINet+~\cite{guo2020convolutional}, another widely recognized benchmark in CSI compression literature. To ensure a fair comparison, we use identical datasets and training protocols across all evaluated models.

We implement two variants of the ADJSCC baseline: (1) the original architecture as presented in~\cite{xu2022deep}, and (2) a parameter-scaled version that matches the complexity of our proposed RD-JSCC model, by increasing the layers and channel dimensions. Additionally, we examine a supervised variant of RD-JSCC in which both the autoencoder and U-Net components are trained end-to-end with an MSE loss. Our primary contribution, RD-JSCC, which integrates an autoencoder with diffusion-based U-Net training, demonstrates performance improvements of an order of magnitude compared to~\cite{xu2022deep} at equivalent NMSE targets.

\emph{Architectural Differences.} To ensure fair comparison across architectures, we maintain identical encoder-decoder designs for ADJSCC, CSINet+, and RD-JSCC, with key distinctions: (1) ADJSCC includes an analog transformation module that maps spatial-frequency domain CSI to a learned representation before compression, (2) ADJSCC and RD-JSCC incorporate SNR-adaptive feature scaling during encoding and decoding, while CSINet+ does not. Our RD-JSCC framework extends the base autoencoder architecture (ADJSCC without the transformation module) by adding a U-Net-based diffusion denoising module at the decoder. This modular design isolates the contribution of diffusion-based refinement from other architectural choices.

Although parameter matching via layer and dimension adjustments may not represent optimal baseline configurations, the order of magnitude performance difference when our approach outperforms the best non-diffusion baseline, as seen in Fig.~\ref{fig:results_cost2100}, highlights fundamental limitations of existing supervised learning methods for joint source channel coding of CSI in complex channels. These substantial gains, even under capacity-constrained conditions, validate our architectural innovations beyond parameter-count considerations.



\subsection{Dataset}\label{sec:dataset}

We evaluate our method on the COST2100 outdoor dataset~\cite{liu2012cost}, which is widely used for its realistic and complex channel characteristics. The dataset provides $10^5$ training samples and $2 \times 10^4$ test samples. It is important to note that the original spatial-frequency domain representations are not publicly available; instead, we utilize the provided $32 \times 32$ cropped complex channel matrices in the angular-delay domain. Consequently, we omit the non-linear transformation module used in ADJSCC and retain only the inner encoder-decoder components that operate directly on the $32 \times 32$ angular-delay domain complex inputs. Since the input dimensions of the inner module match those of the COST2100 cropped channels, this setup enables a fair and consistent comparison. Following standard preprocessing practices, we apply min-max normalization to scale the complex channel matrices to $[0, 1]$ by dividing by twice the maximum absolute value and adding 0.5. All evaluations are reported in terms of NMSE measured in the angular-delay domain.

\begin{figure}[t]
    \centering
 	\includegraphics[width=1.0\linewidth]{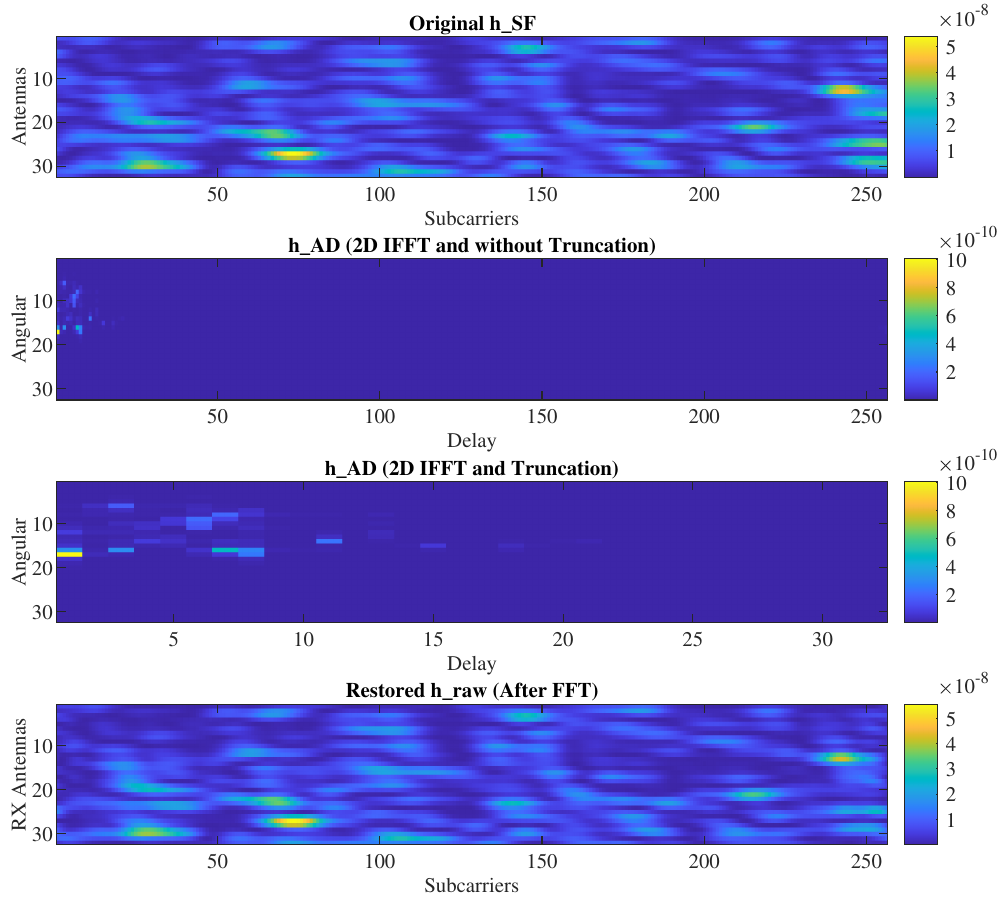}
 	\captionsetup{font=small}
 	\caption{ CSI preprocessing for 3GPP Indoor dataset. The spatial-frequency channel undergoes 2D IFFT transformation to the angular-delay domain, revealing sparsity that enables truncation to $32 \times 32$ dominant coefficients. Reconstruction via 2D FFT achieves near-lossless recovery of the original channel.}
 	\label{fig:csi_viz}
\end{figure}

Since there is no associated uplink channel for the COST2100 outdoor dataset, we generate uplink channel realizations using QuaDRiGa~\cite{jaeckel2017quadriga}, adhering to the 3GPP TR 38.901 channel specification~\cite{tr385g}, with an uplink carrier frequency of $5.4~\text{GHz}$. We adopt the simulation configuration outlined in~\cite{xu2022deep} and assume a dominant line-of-sight (LOS) component. The base station (BS) is positioned at the center of a $20~\text{m} \times 20~\text{m}$ area and is equipped with a uniform linear array (ULA) comprising $N_{\mathrm{t}} = 32$ omnidirectional elements spaced at half-wavelength intervals. The user equipment (UE) employs a single omnidirectional antenna ($N_{\mathrm{r}} = 1$). The antenna heights are set to $3\,\text{m}$ at the BS and $1.5\,\text{m}$ at the UE. The simulation includes $N_c = 32$ subcarriers in the uplink transmission. A sampled channel realization in SF and AD domains for the 3GPP Indoor dataset is provided in Fig.~\ref{fig:csi_viz}, to explicitly show the sparse nature of the channel in the AD domain and the ability to truncate the channel without losing any significant information in the SF domain.

\subsection{Model configuration}

\textit{Encoder.\ } To ensure a low-complexity design for the UE, we choose a small number of convolution channels and do not utilize any residual connections in the encoder architecture. The detailed architectural choices are summarized in~\tabref{tab:enc_arch}.

\begin{table}[!htb]
  \centering
  \begin{tabular}{l c c }
    \toprule
    Layer
      & \makecell[c]{Channels} 
      & \makecell[c]{Kernel size} \\
    \midrule
    Input Conv     & 2    & (11,11) \\
    Conv layer 1    & 32     & (9,9) \\
    Conv layer 2     & 48    & (7,7) \\
    Output Conv    & 2     & (5,5) \\

        \bottomrule
  \end{tabular}
  \caption{CNN encoder.}
  \label{tab:enc_arch}
\end{table}

\textit{Decoder.\ }The receiver employs a series of residual blocks with varying channel widths and kernel sizes. This deeper architecture, combined with residual connections, enhances reconstruction performance while introducing only a modest increase in decoding latency. The detailed architectural choices are summarized in~\tabref{tab:dec_arch}.

\begin{table}[!htb]
  \centering
  \begin{tabular}{l c c }
    \toprule
    Layer
      & \makecell[c]{Channels} 
      & \makecell[c]{Kernel size} \\
    \midrule
    Input Conv      & 2     & (7,7) \\
    Residual Block Conv layer 1     & 16    & (7,7) \\
    Residual Block Conv layer 2     & 24    & (5,5) \\
    Residual Block Conv layer 3     & 2    & (3,3) \\

        \bottomrule
  \end{tabular}
  \caption{Residual Block decoder.}
  \label{tab:dec_arch}
\end{table}

\textit{Diffusion.\ }The architecture for the diffusion denoising network is based on~\cite{yang2023lossy} and~\cite{kim2025generative}. We choose an initial embedding width of $64$ channels and dimension multipliers $\{1,\,2,\,3,\,4\}$ for the successive down-sampling and mirrored up-sampling stages.  
Skip connections link each down-sampling block to its symmetric up-sampling counterpart, preserving high-resolution features throughout the reverse diffusion trajectory.  
The U-Net outputs a refined estimate of the CSI matrix that is fed back, iteratively refining the coarse estimate, based on the number of steps used in reverse diffusion.

\subsection{Hyperparameters for training}\label{sec:hyper}
To train the diffusion model, we adopt a similar training methodology and set of hyperparameters used in~\cite{kim2025generative}. 
The Adam optimizer is employed with a cosine annealing learning rate scheduler that goes from an initial learning rate of $3 \times 10^{-4}$ to $1 \times 10^{-5}$, and a batch size of $100$ is used.  For the diffusion process, we use a cosine beta schedule for determining the noise variance at each step.


As described in~\secref{sec:resdiff}, the model is trained in two stages. In the first stage, the autoencoder model is trained for $N_\text{train} = 10^5$ iterations using the MSE loss~\eqref{eq:loss_mse}. In the second stage, the diffusion-based U-Net is trained for $N_\text{train} = 10^6$ iterations. The coarse estimate produced by the autoencoder serves as the initialization point for reverse diffusion, with denoising performed over $T=20$ steps during training. However, during inference, we can use a 2-step denoising approach for reverse diffusion, with a small performance penalty. This is primarily enabled by training the diffusion model using \textit{x-prediction} instead of Gaussian noise, thus making the inference $10 \times$ faster. Performance is evaluated using the NMSE, which is given as $\mathbb{E} \left[ {\|\mathbf{z} - \hat{\mathbf{z}} \|}/{\|\mathbf{z}\|} \right]$ for the ground truth $\mathbf{z}$ and reconstruction $\hat{\mathbf{z}}$.

Further, regarding the noise scheduling, our derivation in Equation~\eqref{eq:z_t_1} adopts deterministic DDIM sampling ($\sigma_t = 0$) and a linear residual trajectory assumption. This design is motivated by two key factors for CSI feedback:

\begin{itemize}
    \item \textit{Reconstruction Consistency:} Unlike generative tasks, CSI recovery requires high fidelity. Deterministic sampling eliminates stochastic variance, ensuring a unique, optimal reconstruction for each compressed latent.
    \item \textit{Low-Latency Robustness:} In few-step regime required for faster inference, deterministic sampling significantly outperforms stochastic methods (e.g., DDPM) which require many steps ($T \gg 50$) to average out noise \cite{song2020denoising}.
\end{itemize}


\subsection{Results}\label{sec:results_sub}

\begin{figure}[t]
    \centering
    \includegraphics[width=\linewidth]{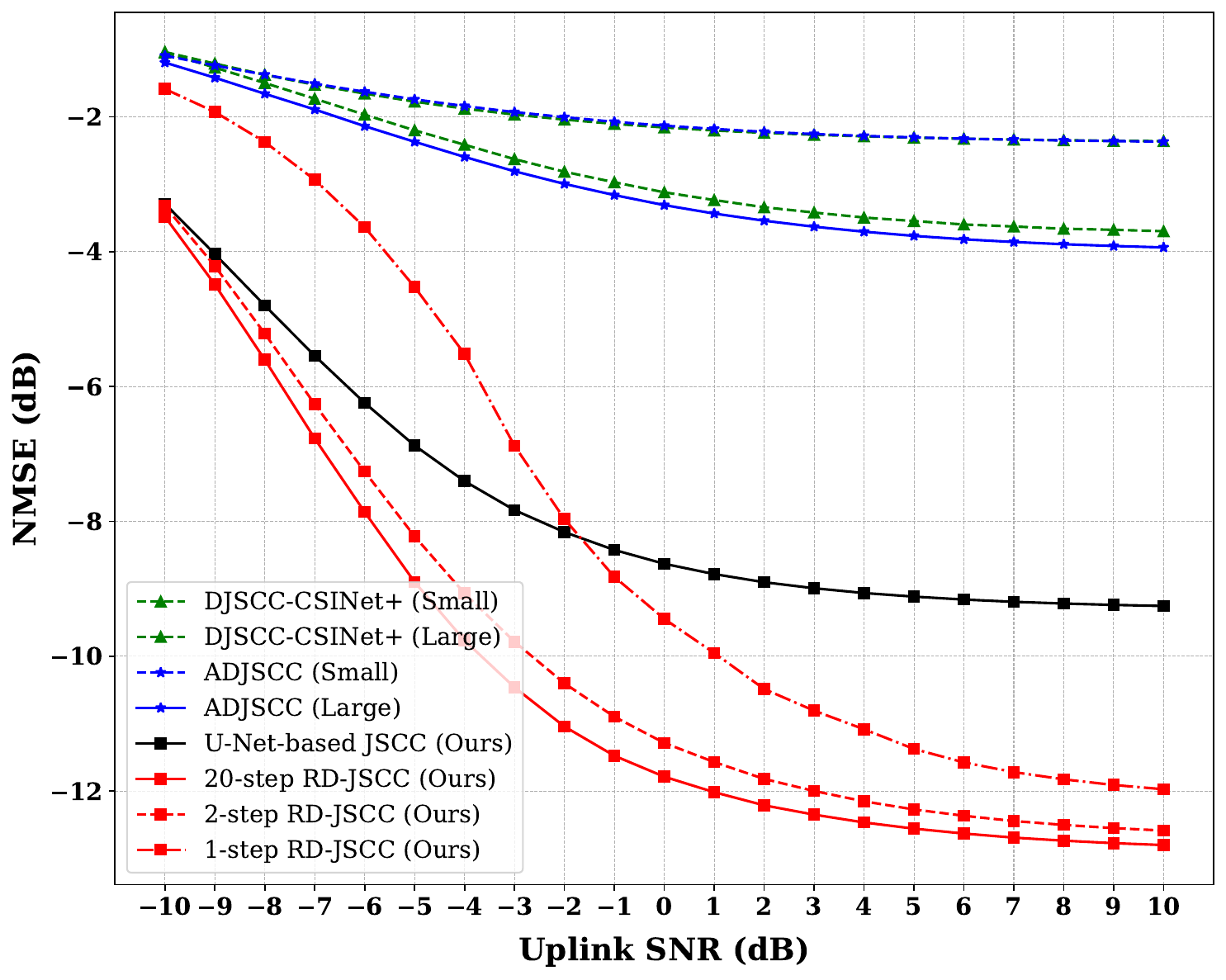}
    \captionsetup{font=small}
    \caption{RD-JSCC achieves an order-of-magnitude improvement over state-of-the-art deep JSCC baselines on the COST2100 outdoor dataset for $k=16$. While traditional supervised learning approaches fail under complex channel distributions, our diffusion-based iterative refinement successfully models the underlying statistics.}
    \label{fig:results_cost2100}
\end{figure}

In~\figref{fig:results_cost2100}, we present the NMSE in reconstruction for the angular-delay domain CSI measurements of the COST2100 outdoor dataset. We configured the feedback bandwidth to $k=16$, compressing each $32 \times 32$ CSI measurement (1024 complex coefficients) to a 16-dimensional complex latent vector, achieving a compression rate of $\sfrac{1}{64}$. Note that for the COST2100 outdoor dataset, the NMSE reported is in the cropped angular-delay domain.

Our evaluation begins with the exact architectures from ADJSCC~\cite{xu2022deep} and CSINet+~\cite{guo2020convolutional}, which demonstrate notably poor performance. When scaling these architectures by incorporating additional intermediate layers and increasing the number of convolutional channels, performance improves only marginally, yet all schemes still saturate at an NMSE greater than -4 dB. This reveals that conventional autoencoder approaches trained with supervised learning objectives exhibit poor scaling characteristics relative to model size and fail to adequately capture the underlying channel characteristics despite substantial parameter counts.

We then developed a hybrid architecture combining the original autoencoder model from~\cite{xu2022deep} with a U-Net backbone for denoising. The U-Net backbone has been widely employed in image processing for enhanced denoising, deblurring, and, more recently, image generation. Our hybrid model, termed U-Net-based JSCC, was trained with the same supervised objective as~\cite{xu2022deep}, namely, minimizing end-to-end MSE. Remarkably, despite having a parameter count comparable to that of the larger variants of ADJSCC and CSINet+, the U-Net-based JSCC achieves significant performance improvements, achieving an NMSE below -9 dB. This highlights the importance of architectures that scale effectively with increasing model size.

Finally, we trained the U-Net-based JSCC model with a diffusion objective rather than a supervised objective. This approach, which we refer to as RD-JSCC, achieves the best performance among all evaluated schemes. During decoding at the receiver, the autoencoder head first produces a coarse estimate of the CSI. The U-Net model then initializes reverse diffusion with this coarse estimate and refines the CSI at each denoising step, iteratively feeding the improved estimate back into the U-Net until the specified number of denoising steps is completed. While our default configuration uses 20 denoising steps, we found that a 2-step reverse diffusion minimizes inference latency while incurring only a negligible NMSE penalty, achieving an NMSE below -12 dB. Notably, 1-step diffusion exhibits significant performance degradation.
\begin{remark}
Diffusion-based modeling demonstrates a clear advantage in capturing statistically complex channel distributions that conventional supervised methods struggle to represent effectively. This advantage stems from the diffusion framework's ability to learn high-dimensional probability distributions through iterative refinement.
\end{remark}

\begin{figure}[t]
    \centering
 	\includegraphics[width=1.0\linewidth]{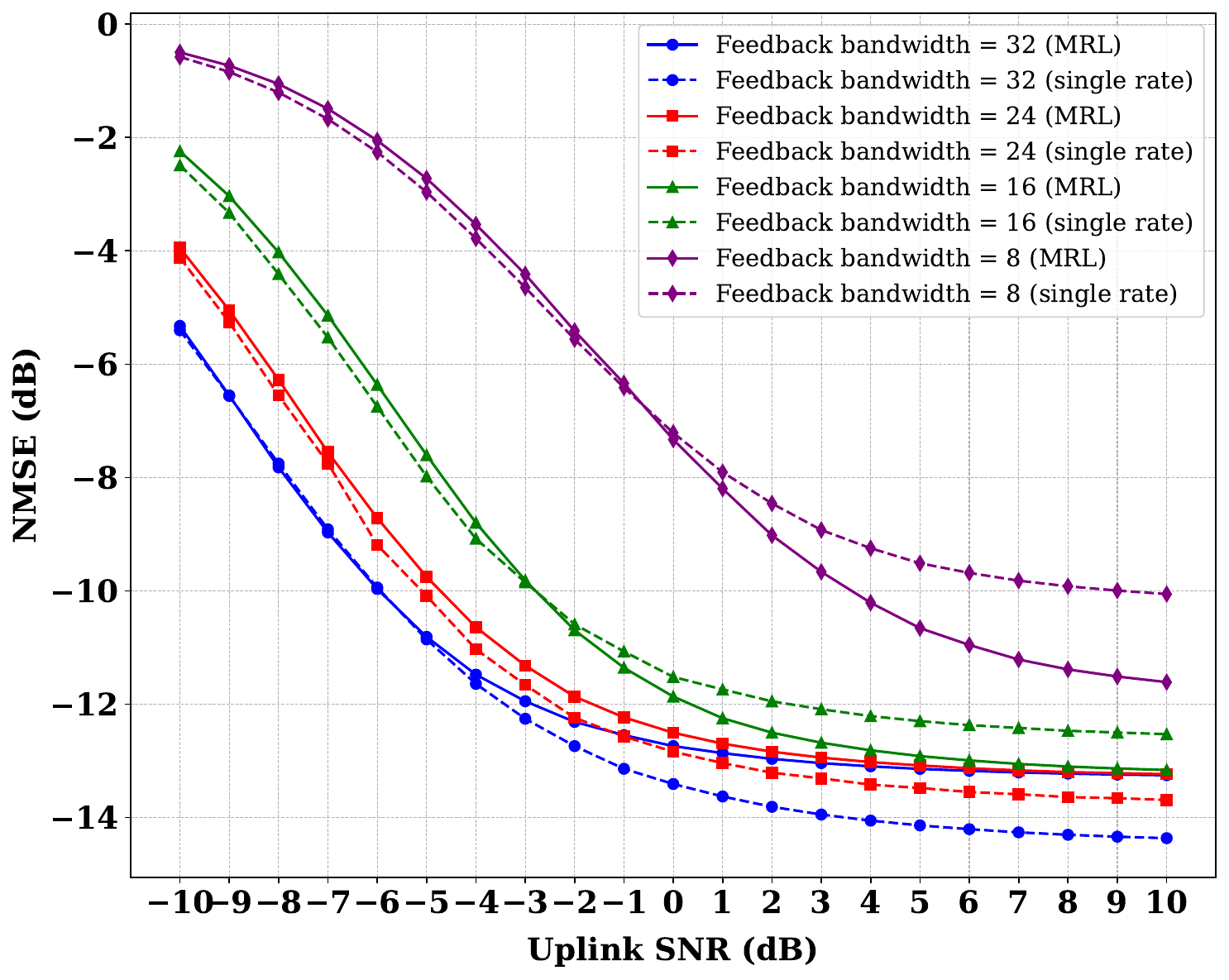}
 	\captionsetup{font=small}
 	\caption{Optimizing for multiple feedback bandwidths: Using a single model with MRL objective, RD-JSCC dynamically adjusts the latent dimension based on available uplink bandwidth, achieving efficient rate-distortion trade-offs across multiple feedback rates.
}
 	\label{fig:mrl}
\end{figure}

\textit{Variable-Rate Compression.}
Further, we now consider the performance of RD-JSCC under the MRL objective to simultaneously optimize the performance for multiple rates, as shown in Fig.~\ref{fig:mrl}. The MRL model exhibits a rate-dependent performance characteristic: at lower bandwidths ($k \leq 16$), it achieves superior reconstruction compared to rate-specific models, while at higher bandwidths ($k \geq 24$), dedicated models show marginal gains. This behavior is expected, as the MRL objective optimizes average performance across the compression-rate spectrum (Eq.~\ref{eq:loss_mrl}), trading slight suboptimality at individual rates for the practical advantages of supporting $25$ rates with a single model and eliminating the storage/switching overhead of maintaining multiple rate-specific architectures.

\subsection{NMSE Performance Translation to BLER}\label{sec:bler}

\begin{figure}[t]
    \centering
 	\includegraphics[width=1.0\linewidth]{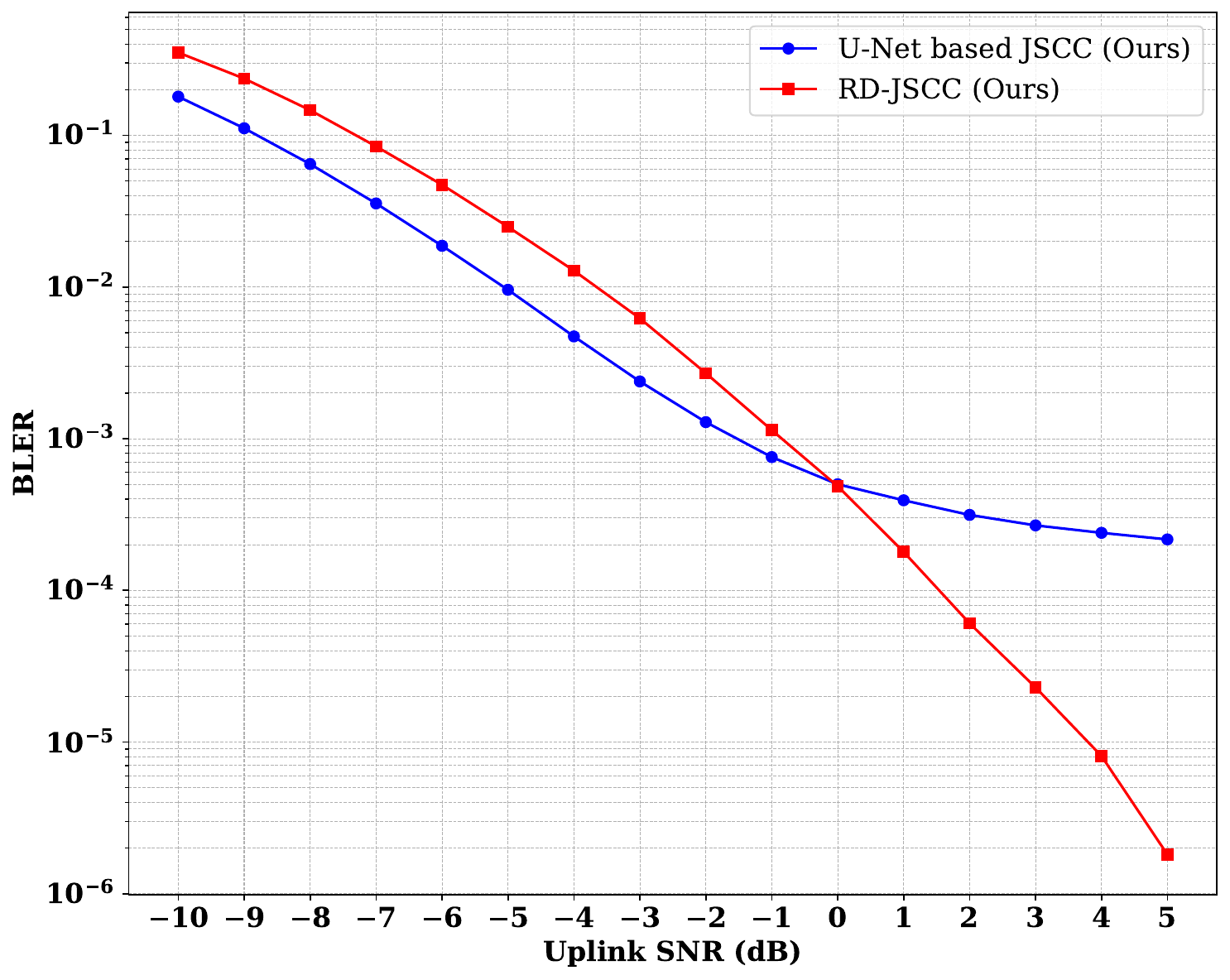}
 	\captionsetup{font=small}
 	\caption{Uncoded BLER comparison for COST2100 shows that RD-JSCC continues improving at high SNR while U-Net-based JSCC saturates. The diffusion-based approach achieves BLER $\approx 10^{-6}$ at 5 dB, demonstrating two orders of magnitude improvement over the U-Net's saturation at $10^{-4}$.}
 	\label{fig:bler}
\end{figure}

While NMSE is widely used to report CSI compression performance, block error rate (BLER) provides a more direct measure of practical system reliability. We therefore evaluate the uncoded BLER performance to assess how NMSE improvements translate to error rate reductions. For the COST2100 dataset, we construct SF-domain data from the available AD-domain measurements using the fast Fourier transform (FFT), assuming a lossless SF-to-AD transformation in the original dataset. As shown in \figref{fig:bler}, both methods achieve similar BLER performance at low SNR values. However, in the high-SNR regime, the U-Net-based approach saturates around $10^{-4}$ BLER, while the diffusion-based RD-JSCC continues improving and reaches BLER $\approx 10^{-6}$ at 5 dB uplink SNR. This performance difference correlates with the underlying reconstruction quality: U-Net-based JSCC achieves NMSE $\approx -9.2$ dB while RD-JSCC achieves NMSE $\approx -12.2$ dB, at an SNR of 5 dB in the uplink. It is interesting to note that the diffusion-based approach underperforms at low SNRs because it relies on the inductive bias of the learned channel distribution rather than the heavily corrupted initial estimate, generating statistically plausible but potentially inaccurate reconstructions, a known limitation of diffusion models under high noise conditions. Conversely, the supervised U-Net maintains stable performance even at low SNRs, while diffusion-based refinement excels at high SNRs where sufficiently accurate initial estimates enable effective iterative enhancement.
\begin{remark}
For the COST2100-outdoor dataset, at an uplink SNR of 5 dB, a 3 dB improvement in CSI NMSE corresponds to approximately two orders of magnitude improvement in BLER, demonstrating that NMSE gains from the diffusion-based approach translate into a reduced error floor at high SNR, thereby improving reliability.
\end{remark}

\subsection{Computational Complexity}\label{sec:complexity}

We now evaluate the computational complexity of RD-JSCC by measuring the throughput of each module on both the COST2100 outdoor and 3GPP indoor datasets. Throughput is measured as samples processed per second (samples/s), averaged across multiple runs with a batch size of $10^3$. All simulations were conducted on a system equipped with an \texttt{AMD Ryzen Threadripper PRO 5975WX 32-Core processor} and a single \texttt{NVIDIA GeForce RTX 4090 GPU}. Training of the RD-JSCC model for the COST2100 dataset takes $\approx15$ GPU hours, and the smaller RD-JSCC model trained on the 3GPP indoor dataset takes $\approx 6.8$ GPU hours.

As shown in Table~\ref{tab:throughput}, the encoder and decoder exhibit high inference throughput due to their lightweight convolutional architecture. The encoder consistently achieves throughput near $9.5 \times 10^4$ samples/s across both datasets, making it suitable for low-power UEs. The decoder is slightly more complex due to residual layers, but still maintains a high throughput of approximately $8.25 \times 10^4$ samples/s.

In contrast, the diffusion model, though delivering significant performance improvements under complex channel conditions, introduces a higher computational cost. The 2-step residual diffusion refinement achieves throughput of $3.9 \times 10^3$ samples/s on COST2100 and $8.2 \times 10^3$ samples/s on the 3GPP indoor dataset. For applications requiring higher fidelity, the full 20-step diffusion yields throughput in the $10^2$ samples/s range, although the improvements in NMSE are marginal compared to 2-step diffusion. Hence, invoking the diffusion refinement module at the decoder can incur a throughput penalty of $10\times$ to $100\times$, and should therefore be used judiciously, only when necessary, based on the underlying channel complexity. Table~\ref{tab:throughput} also shows the inference throughput using CPU, which shows $25-30 \times$ degradation in performance.

These results reinforce the practicality of RD-JSCC's hybrid decoding strategy, unlike a single diffusion-based solution presented in~\cite{kim2025generative}. For simple channels, the decoder alone may suffice, whereas in complex scenarios, diffusion-based refinement can be selectively invoked. Furthermore, early-exit and low-step inference modes offer flexible trade-offs between complexity and performance.


\begin{table}[t]
    \centering
    \scriptsize
    \begin{tabular}{lcccc}
        \toprule
        \multirow{2}{*}{\textbf{Module}} & 
        \multicolumn{2}{c}{\textbf{COST2100}} & 
        \multicolumn{2}{c}{\textbf{3GPP Indoor}} \\
        \cmidrule(lr){2-3} \cmidrule(lr){4-5}
        & \textbf{GPU} & \textbf{CPU} & \textbf{GPU} & \textbf{CPU} \\
        \midrule
        Encoder & $9.5 \times 10^4$ & $3.8 \times 10^3$ & $9.5 \times 10^4$ & $3.8 \times 10^3$ \\
        Decoder & $8.2 \times 10^4$ & $3.4 \times 10^3$ & $8.2 \times 10^4$ & $3.4 \times 10^3$ \\
        Diffusion (2-step) & $3.9 \times 10^3$ & $1.3 \times 10^2$ & $8.5 \times 10^3$ & $2.8 \times 10^2$ \\
        Diffusion (20-step) & $3.9 \times 10^2$ & $1.3 \times 10^1$ & $8.5 \times 10^2$ & $2.8 \times 10^1$ \\
        \bottomrule
    \end{tabular}
    \caption{ GPU and CPU throughput (samples/s) of RD-JSCC modules for COST2100 and 3GPP indoor datasets. A lower-complexity channel requires a smaller model, thereby increasing throughput.}
    \label{tab:throughput}
\end{table}

\section{Practical considerations}\label{sec:practical}

\subsection{Channel Distribution vs. Model Complexity}\label{sec:model_choice}

While diffusion models have the potential to significantly enhance CSI reconstruction quality at the base station, they also introduce substantial computational complexity. Therefore, it is crucial to justify this added complexity by employing diffusion-based refinement only when truly necessary. In the current state-of-the-art diffusion-based CSI compression method~\cite{kim2025generative}, the same network architecture is applied across both the Clustered Delay Line (CDL) and the more challenging COST2100 outdoor datasets, despite the latter being considerably more complex. Moreover, the comparisons to existing baselines do not account for the substantial differences in model size. For example, baseline models used in~\cite{kim2025generative}, such as CSINet and CRNet, typically have approximately 400K parameters, whereas the generative diffusion-based approach presented here utilizes approximately 15M parameters. This significant disparity in parameter count complicates assessing whether performance gains arise from the diffusion modeling itself or simply from increased model size. 

In this section, we conduct a systematic ablation study to isolate the benefits of diffusion modeling and to understand its advantages relative to the underlying channel complexity.

\textit{Architecture Choices.\ } We evaluate three different architectures for this study. First, we consider a simple convolutional autoencoder trained in a supervised manner. Second, we consider an autoencoder followed by a U-Net denoising network, which is also trained end-to-end with a supervised loss. Finally, we evaluate the RD-JSCC model, where the autoencoder and U-Net are trained jointly using the residual diffusion objective. In all three cases, the architectures are appropriately scaled to maintain comparable parameter counts across models. Complete details of parameters used in each experiment are provided in~\tabref{tab:param_cost2100} and~\tabref{tab:param_3gpp}.

\textit{Channel Model Choices.\ } In addition to the COST2100 outdoor scenario~\cite{liu2012cost} analyzed in~\secref{sec:results}, we now evaluate performance under the indoor open-area channel model specified in 3GPP TR 38.901~\cite{tr385g}. This environment is characterized by significantly lower spatial complexity and information density than COST2100, providing a useful contrast in dataset complexity. As seen in~\secref{sec:results}, we first convert the spatial-frequency data into the cropped angular-delay domain, preserving the first $32$ subcarriers. But, after decompression, we convert the estimated channel matrices from the cropped angular-delay domain back to the spatial-frequency domain and subsequently present the final NMSE results in the original SF domain.

\begin{figure}[t]
    \centering
    \includegraphics[width=\linewidth]{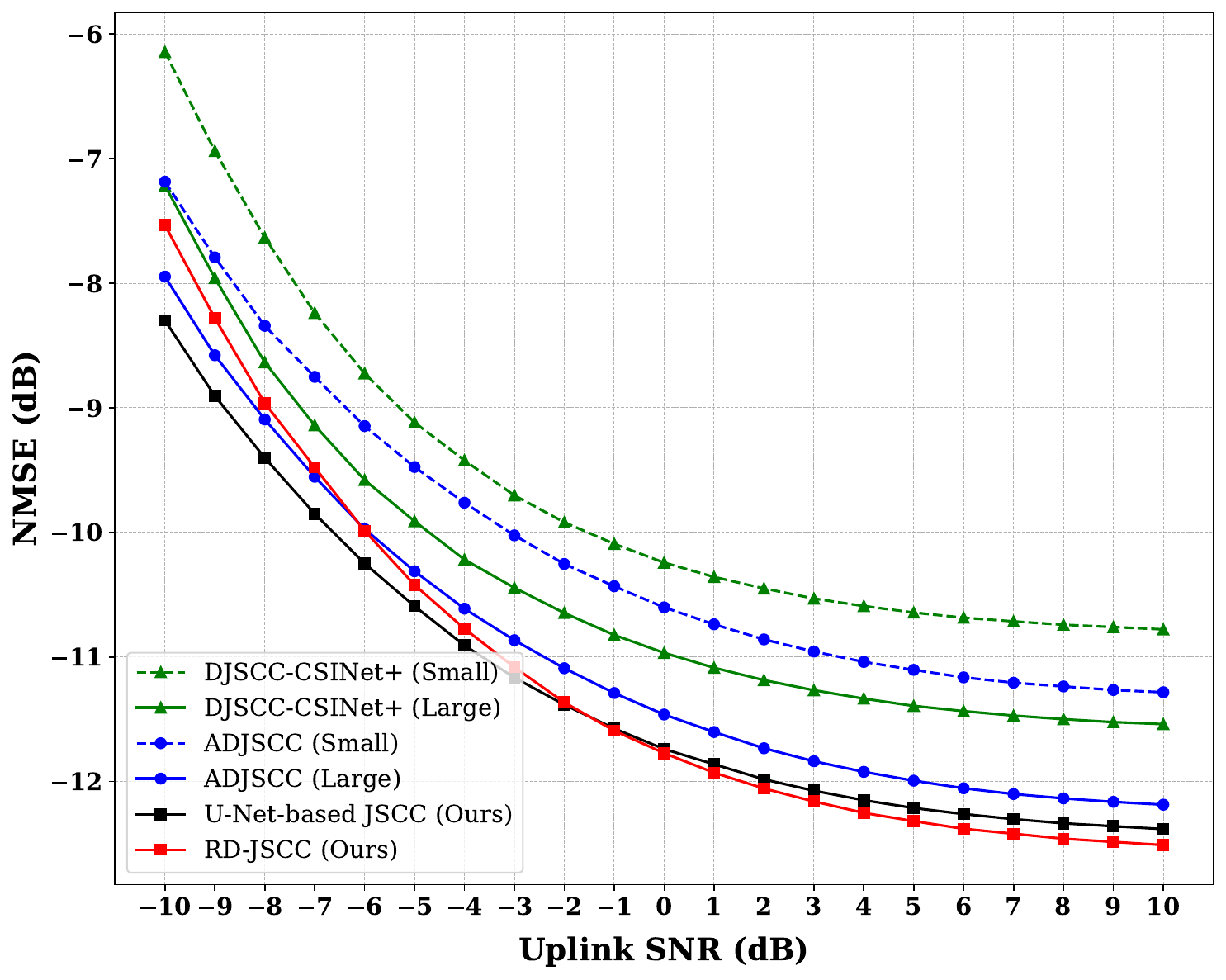}
    \captionsetup{font=small}
    \caption{Under low-complexity channel conditions, such as the 3GPP indoor scenario, the performance gap between diffusion-based and autoencoder-based approaches narrows significantly.}
    \label{fig:results_indoor}
\end{figure}

Analysis of results in~\figref{fig:results_indoor} reveals that the performance gap between our diffusion-based approach and autoencoder baselines narrows considerably for the 3GPP indoor dataset, in stark contrast to the COST2100 outdoor results presented in~\figref{fig:results_cost2100}. 
This observation highlights an important nuance: 

\begin{remark}
Despite the general superiority of diffusion-based CSI compression, the complexity-performance trade-off becomes less favorable when modeling simpler channel environments, where conventional low-complexity autoencoder architectures may offer a more efficient solution.
\end{remark}

 Notably, the flexible formulation of our residual diffusion framework provides an additional operational advantage: it allows early termination of the decoding process after stage 1, regardless of prevailing channel conditions, thereby enabling adaptive decoding-complexity scaling based on application requirements.

\begin{table}[t]
  \centering
  \renewcommand{\arraystretch}{1.05}
  \setlength{\tabcolsep}{3.5pt}
  \scriptsize
  \resizebox{0.92\linewidth}{!}{
    \begin{tabular}{l cc cc cc}
      \toprule
      \rowcolor{gray!12}
      & \multicolumn{2}{c}{\textbf{Encoder}} 
      & \multicolumn{2}{c}{\textbf{Decoder}} 
      & \multicolumn{2}{c}{\textbf{Denoiser}} \\
      \cmidrule(lr){2-3} \cmidrule(lr){4-5} \cmidrule(lr){6-7}
      \rowcolor{gray!8}
      \textbf{Model} & Params & FLOPs & Params & FLOPs & Params & FLOPs \\
      \midrule
      DJSCC-CSINet+ (Small) & 152K & 21.3M & 118K & 26.4M & -- & -- \\
      DJSCC-CSINet+ (Large) & 152K & 21.3M & 12M  & 320M & -- & -- \\
      ADJSCC (Small)        & 152K & 21.3M & 118K & 26.4M & -- & -- \\
      ADJSCC (Large)        & 152K & 21.3M & 12M  & 320M & -- & -- \\
      \specialrule{.1em}{.05em}{.05em}
      U-Net-based JSCC      & 152K & 21.3M & 118K & 26.4M & 13.7M & 135.6M \\
      RD-JSCC (2-step)      & 152K & 21.3M & 118K & 26.4M & 13.9M & 271.8M \\
      \bottomrule
    \end{tabular}
  }
  \caption{Parameter counts and FLOPs comparison across methods for COST2100 dataset.}
  \label{tab:param_cost2100}
\end{table}

\begin{table}[!htb]
  \centering
  \renewcommand{\arraystretch}{1.1}
  \setlength{\tabcolsep}{4pt}
  \scriptsize
  \resizebox{0.92\linewidth}{!}{
    \begin{tabular}{l cc cc cc}
      \toprule
      \rowcolor{gray!12}
      & \multicolumn{2}{c}{\textbf{Encoder}} 
      & \multicolumn{2}{c}{\textbf{Decoder}} 
      & \multicolumn{2}{c}{\textbf{Denoiser}} \\
      \cmidrule(lr){2-3} \cmidrule(lr){4-5} \cmidrule(lr){6-7}
      \rowcolor{gray!8}
      \textbf{Model} & Params & FLOPs & Params & FLOPs & Params & FLOPs \\
      \midrule
      DJSCC-CSINet+ (Small) & 167K & 27.8M & 191K & 35.7M & -- & -- \\
      DJSCC-CSINet+ (Large) & 167K & 27.8M & 1.3M & 63.4M & -- & -- \\
      ADJSCC (Small)        & 168K & 28.1M & 197K & 35.9M & -- & -- \\
      ADJSCC (Large)        & 168K & 28.1M & 1.3M & 63.7M & -- & -- \\
      \specialrule{.1em}{.05em}{.05em}
      U-Net-based JSCC      & 152K & 21.3M & 118K & 26.4M & 1.25M & 29.5M \\
      RD-JSCC (2-step)              & 152K & 21.3M & 118K & 26.4M & 1.28M & 59.7M \\
      \bottomrule
    \end{tabular}
  }
  \caption{Parameter counts and FLOPs comparison across methods for 3GPP Indoor dataset.}
  \label{tab:param_3gpp}
\end{table}

\subsection{Imperfect Channel Estimation}\label{sec:che}

In the CSI compression literature, it is common to assume perfect CSI at UE. While this assumption is convenient for algorithm development and benchmarking, it rarely holds true in practical wireless communication systems. In realistic deployments, CSI must be acquired via pilot-based estimation at the receiver, which introduces errors arising from noise, interference, and limited pilot resources. These estimation inaccuracies can significantly degrade the performance of both traditional and learning-based CSI compression schemes.

To bridge this gap, we explicitly incorporate channel estimation errors into our problem formulation. Specifically, we model the estimated downlink channel \( \widetilde{H}_d \) as
\begin{equation}
    \widetilde{H}_d = H_d + E,
\end{equation}
where \( H_d \in \mathbb{C}^{N_c \times N_t} \) is the true downlink channel matrix, and \( E \in \mathbb{C}^{N_c \times N_t} \) represents the estimation error. Each entry of \( E \) is modeled as a zero-mean complex Gaussian random variable, with variance determined by the channel-to-noise ratio (CNR). This additive error model is commonly used to model the uncertainty inherent in practical CSI acquisition~\cite{xu2022deep}.

\begin{figure}[t]
    \centering
 	\includegraphics[width=1.0\linewidth]{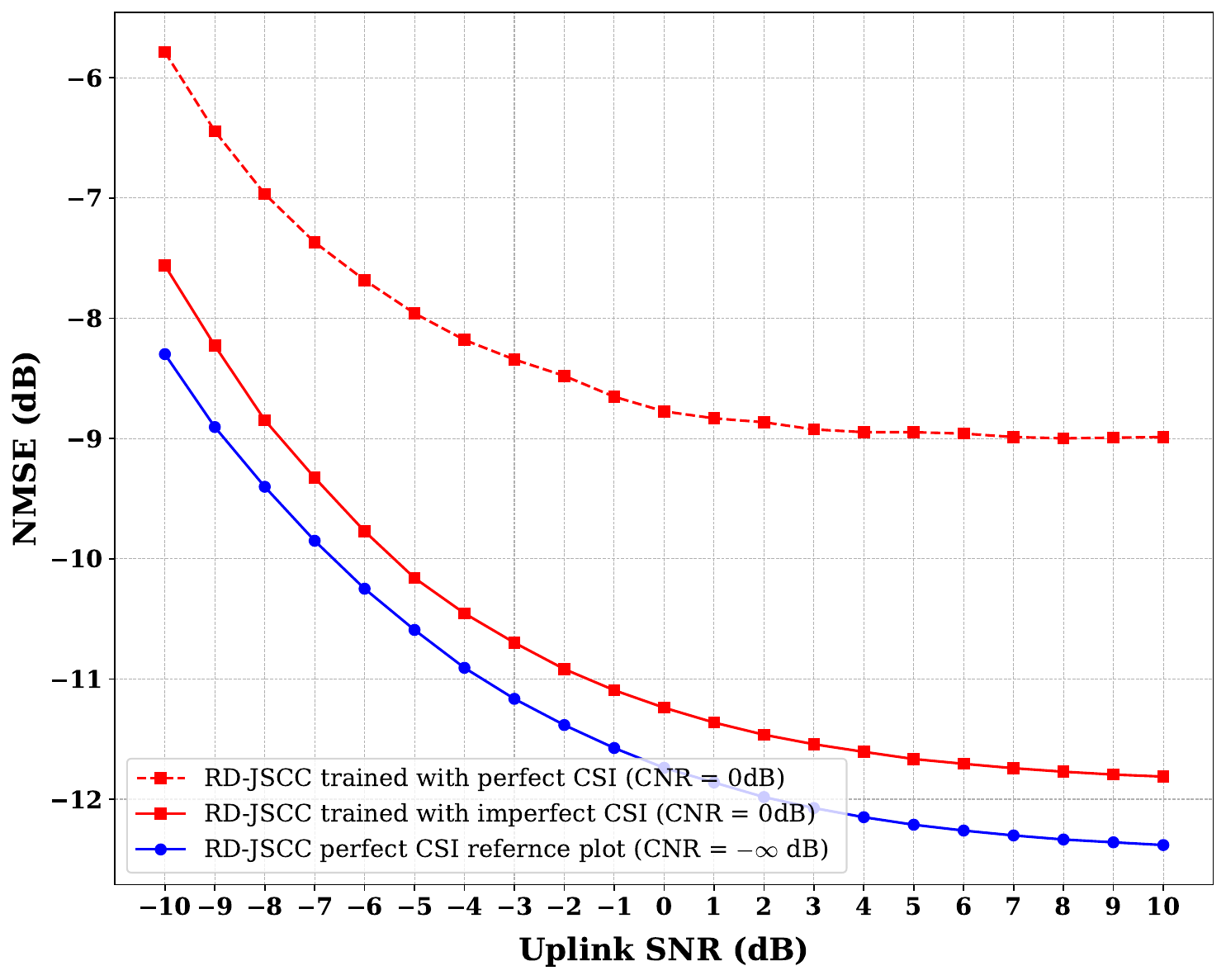}
 	\captionsetup{font=small}
 	\caption{By injecting channel estimation errors into the training pipeline, RD-JSCC becomes robust to imperfect CSI and recovers much of the performance degradation caused by estimation noise when tested at a CNR of $0$ dB.
}
 	\label{fig:imperfect_che}
\end{figure}

In~\figref{fig:imperfect_che}, we observe that RD-JSCC trained using error-free CSI under the perfect CSI assumption suffers a substantial performance drop when exposed to severe channel estimation errors, particularly at a CNR of $0$~dB. However, by incorporating channel estimation error directly into the training pipeline, RD-JSCC learns to operate robustly under imperfect CSI, recovering most of the performance loss.

\subsection{Effect of Quantization}\label{sec:quant}

When using autoencoders for data compression, the encoder typically produces a continuous latent vector $\mathbf{s} \in \mathbb{C}^k$, which is often stored using 32-bit floating-point precision. This imposes significant memory and communication overhead, particularly in resource-constrained settings, such as edge devices and embedded wireless systems. To reduce this burden, we use a companding-based quantization scheme that enables fixed-bit representation of latent vectors with minimal performance degradation. Note that we quantize the real and imaginary values independently, using identical techniques. 

Specifically, we first apply $\mu$-law companding to the latent vector to non-linearly compress its dynamic range, using the transformation:
\[
s'_i = \operatorname{sign}(s_i) \cdot \frac{\log(1 + \mu |s_i|)}{\log(1 + \mu)},
\]
where $\mu$ is a positive scalar hyperparameter (set to $\mu=50$ in our experiments). This transformation increases the quantization resolution near zero, where the latent values are often concentrated. The companded latent $\mathbf{s}'$ is then passed through a uniform scalar quantizer with $B$ bits per dimension. During training, this quantization is implemented with a straight-through estimator to enable end-to-end gradient flow.

\begin{figure}[t]
    \centering
 	\includegraphics[width=1.0\linewidth]{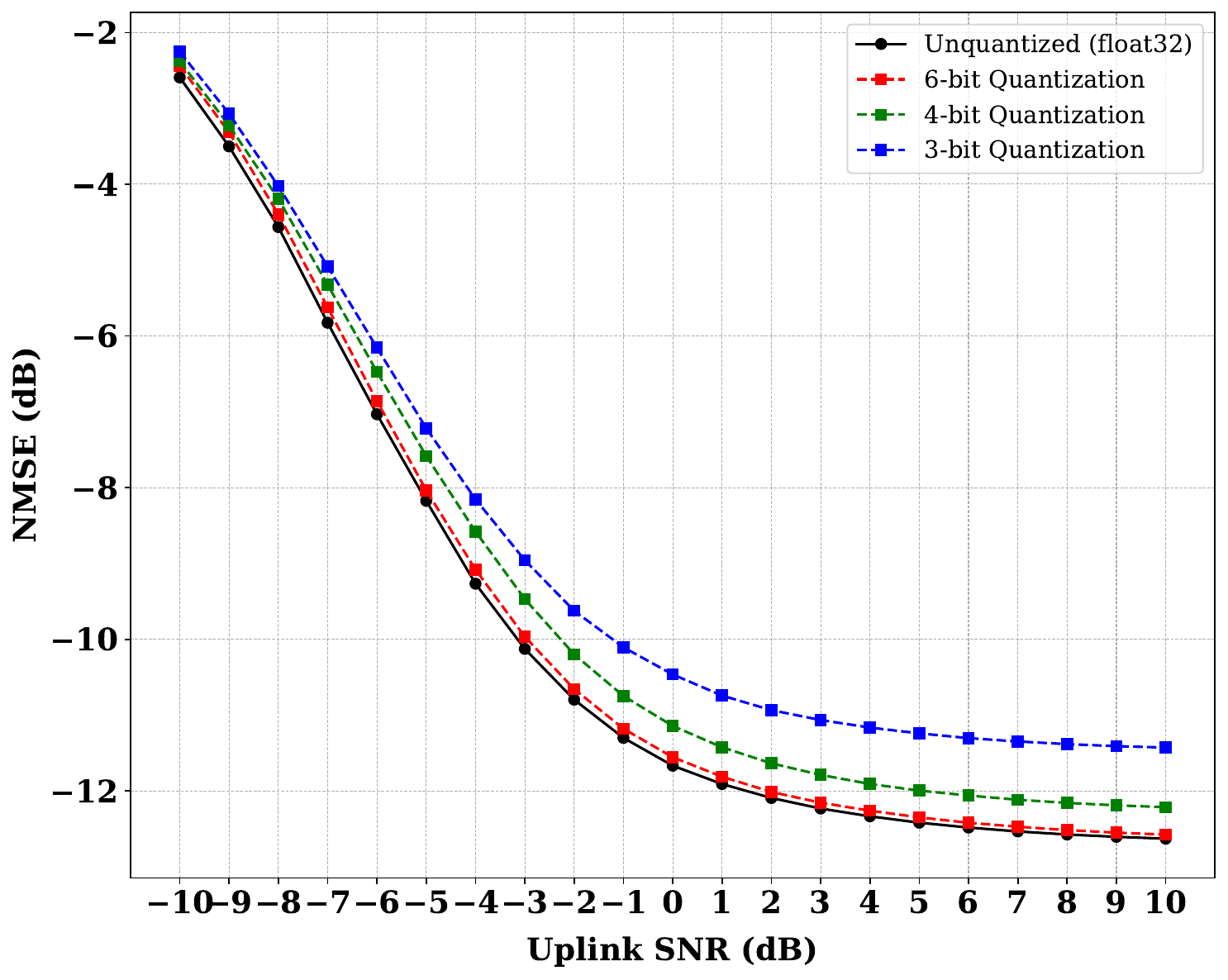}
 	\captionsetup{font=small}
 	\caption{Effect of fixed-bit quantization on reconstruction quality. The encoder output can be quantized to as few as 4 bits per dimension with negligible degradation in reconstruction performance.   }
 	\label{fig:quant}
\end{figure}

At inference, the quantized codes are de-quantized and passed through the inverse $\mu$-law transformation:
\[
\hat{s}_i = \operatorname{sign}(s'_i) \cdot \frac{(1 + \mu)^{|s'_i|} - 1}{\mu},
\]
which approximately reconstructs the original latent vector. As shown in~\figref{fig:quant}, this scheme enables us to quantize the latent space down to $B=4$ bits per dimension with negligible performance loss, offering a practical and efficient solution for deployment in memory and bandwidth-limited systems.

\section{Ablation}\label{sec:ablation}

\subsection{Contribution of Residual Diffusion Formulation}\label{sec:ablation_res}
To quantify the impact of the residual formulation in our denoising diffusion process, we conduct an ablation study comparing our proposed RD-JSCC scheme against a conventional Gaussian Diffusion (GD) JSCC baseline. The baseline follows the standard formulation introduced in~\cite{kim2025generative}, where the reverse diffusion process is initialized from pure Gaussian noise and trained to generate the CSI purely based on the conditioning signal. In contrast, RD-JSCC initiates reverse diffusion with a coarse reconstruction from a lightweight autoencoder and iteratively denoises the residual between the ground-truth CSI and this initial estimate.

\begin{figure}[t]
    \centering
    \includegraphics[width=\linewidth]{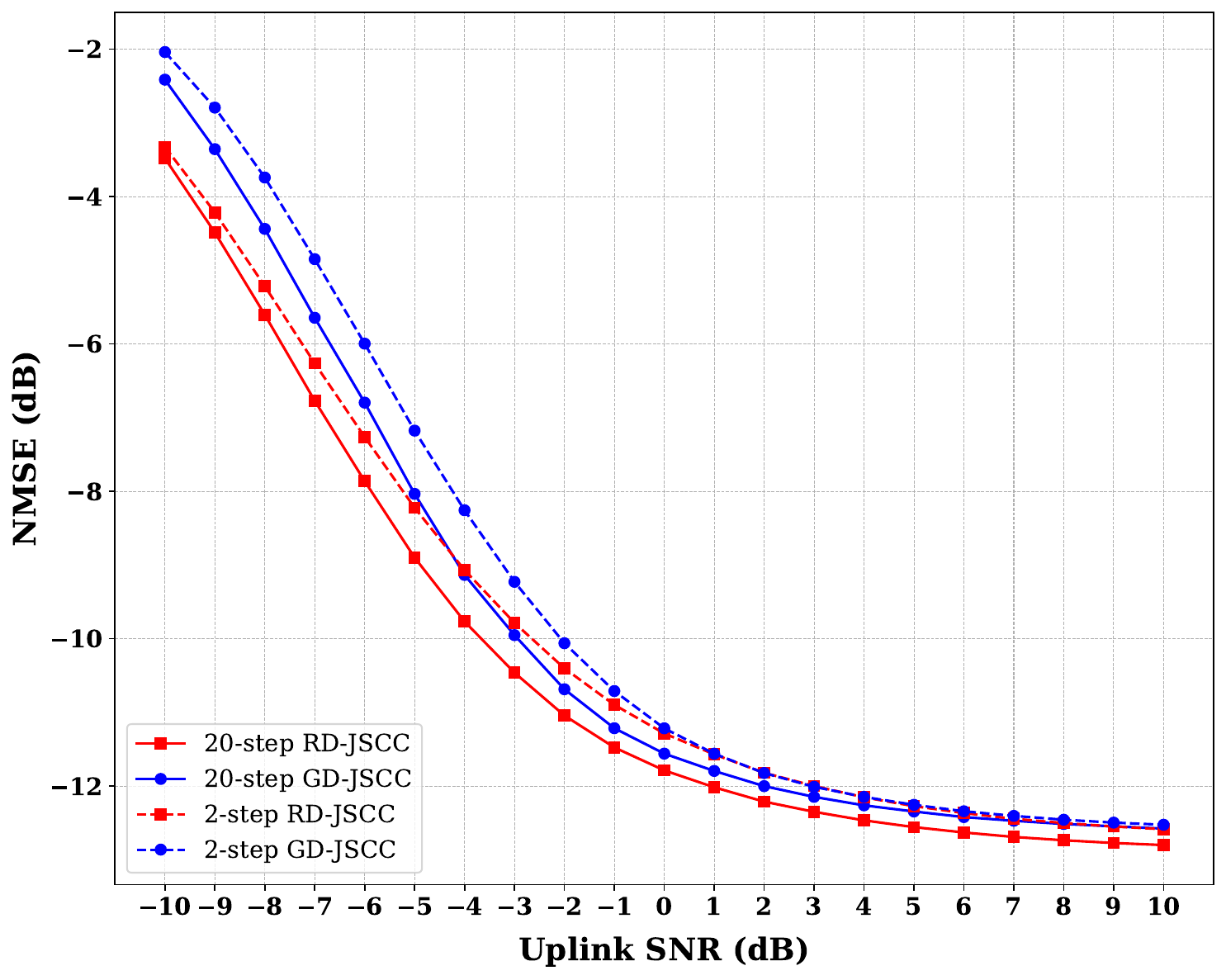}
    \captionsetup{font=small}
    \caption{Comparison between standard GD-JSCC and our proposed RD-JSCC for feedback bandwidth $k=16$. By initializing reverse diffusion with a coarse CSI estimate and modifying the denoising objective to predict residual noise, RD-JSCC achieves up to 1 dB SNR improvement for a target NMSE.}
    \label{fig:ablation_rd}
\end{figure}

This residual formulation enables a modified diffusion objective that focuses on learning the residual signal, which is often sparser and easier to model. As shown in Fig.~\ref{fig:ablation_rd}, RD-JSCC consistently outperforms the GD-JSCC baseline across all tested SNR levels. Notably, it achieves up to a 1~dB improvement in effective SNR for a target NMSE, clearly demonstrating the advantages of the residual formulation in diffusion-based compression.


\subsection{Contribution of Encoder}\label{sec:ablation_enc}
\begin{figure}[t]
    \centering
    \includegraphics[width=\linewidth]{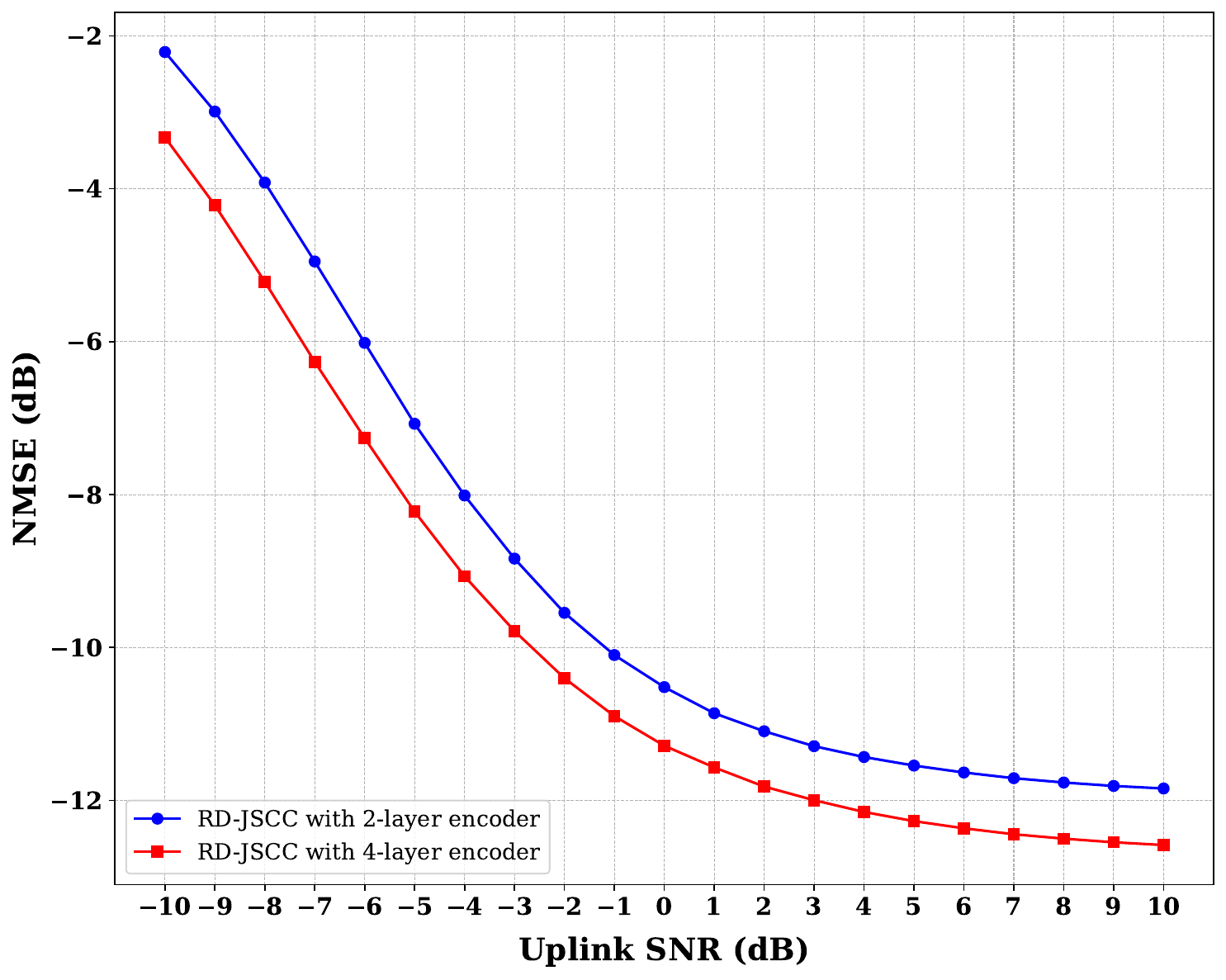}
    \captionsetup{font=small}
    \caption{Impact of encoder complexity on reconstruction performance. Reducing from a 4-layer to a 2-layer architecture with fewer channels degrades NMSE by up to 1 dB on the COST2100 dataset at a feedback bandwidth of $k=16$, despite the encoder's minimal contribution to overall model complexity.}
    \label{fig:ablation_enc}
\end{figure}

To investigate the role of encoder complexity in our architecture, we conducted ablation studies varying the encoder depth and channel capacity. Despite comprising a relatively small fraction of the overall model parameters,~\figref{fig:ablation_enc} demonstrates that encoder design significantly impacts reconstruction quality. When reducing our standard 4-layer encoder to a lighter 2-layer variant with fewer channels, performance deteriorates substantially, by up to 1 dB loss in NMSE at equivalent compression rates. This confirms that the encoder, while computationally efficient, serves a critical function in extracting and preserving essential features from the input signal. These preserved features subsequently enable both the decoder and diffusion modules to achieve high-quality reconstruction. 

The exact architecture choices for the 2-layer and 4-layer encoder are provided in~\tabref{tab:enc_arch_comparison}. The 2-layer encoder has a total of $66$K parameters, and the 4-layer encoder has a total of $152$K parameters. For reference, we recall from~\tabref{tab:param_cost2100} that the full RD-JSCC model has $\approx 14$M parameters. Our findings suggest that appropriate encoder complexity represents an important design consideration that disproportionately influences overall system performance relative to its parameter count.

\begin{table}[!htb]
  \centering
  \begin{tabular}{l c c | c c}
    \toprule
    \multirow{2}{*}{Layer} & 
    \multicolumn{2}{c|}{4-Layer Encoder} & 
    \multicolumn{2}{c}{2-Layer Encoder} \\
    \cmidrule{2-5}
    & Channels & Kernel size & Channels & Kernel size \\
    \midrule
    Input Conv     & 2    & (11,11) & 2 & (7,7) \\
    Conv layer 1    & 32     & (9,9) &  -- & -- \\
    Conv layer 2     & 48    & (7,7) & -- & -- \\
    Output Conv    & 2     & (5,5) & 2 & (7,7) \\
    \bottomrule
  \end{tabular}
  \captionsetup{font=small}
  \caption{Comparison between 4-layer encoder and lightweight 2-layer encoder architectures used in ablation studies. The 4-layer configuration provides superior feature extraction capabilities, leading to better reconstruction performance as demonstrated in Fig.~\ref{fig:ablation_enc}.}
  \label{tab:enc_arch_comparison}
\end{table}

\subsection{Contribution of SNR Adaptation}\label{sec:snr}

We investigate the contribution of SNR side information by training RD-JSCC without the SNR adaptation module and comparing performance across uplink SNR conditions. We see from Fig.~\ref{fig:ablation_enc} that SNR conditioning provides consistent improvements across all SNR values, with up to $0.35$ dB NMSE gain in the high SNR regime and up to $0.1$ dB gain in the lower-SNR region, demonstrating that SNR-aware denoising enables the diffusion model to adapt its reconstruction strategy based on channel quality rather than learning a single solution.

\begin{figure}[t]
    \centering
    \includegraphics[width=\linewidth]{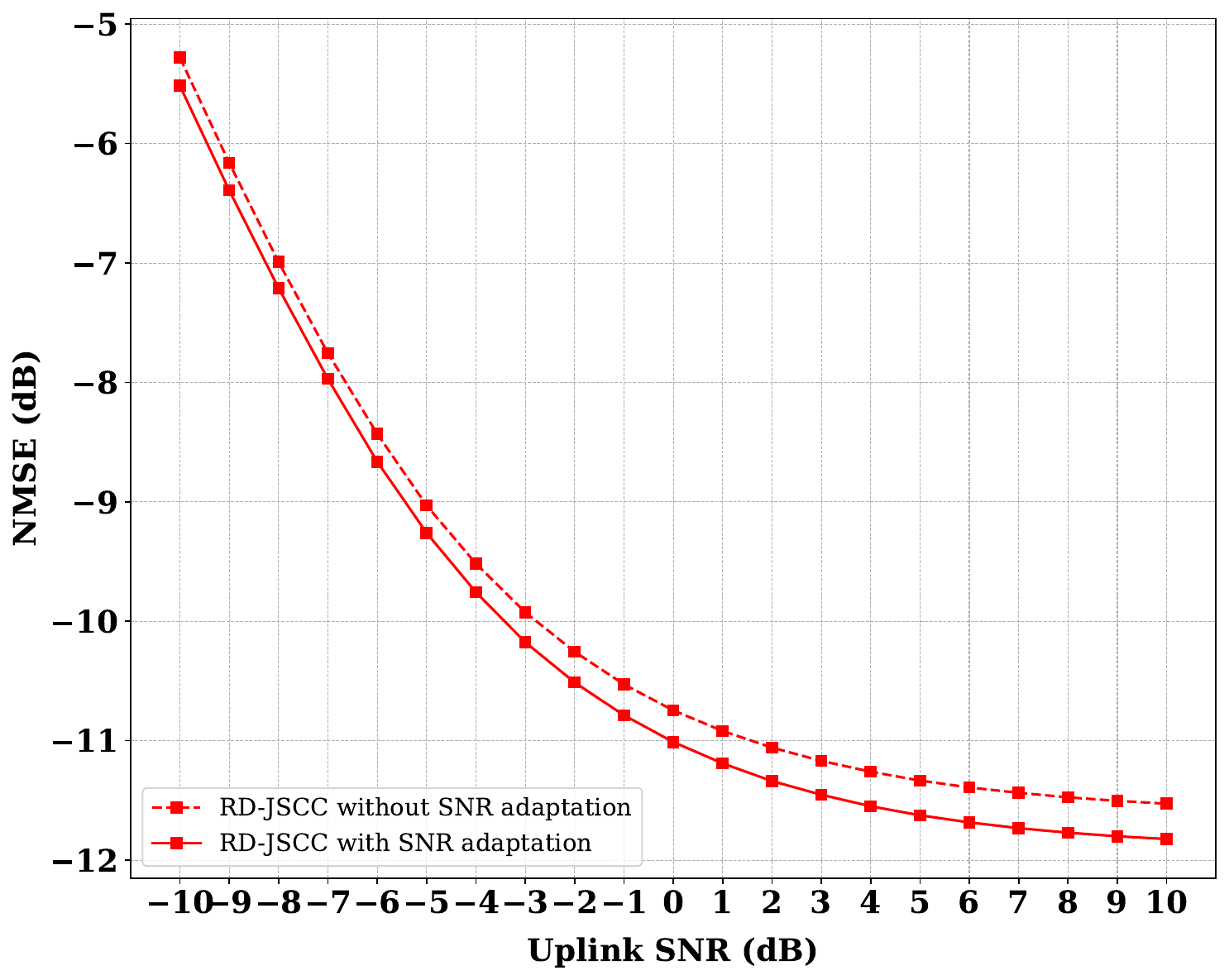}
    \captionsetup{font=small}
    \caption{Impact of SNR side-information on RD-JSCC performance. Removing the SNR adaptation module results in an NMSE degradation of up to $0.35$ dB for a given SNR.}
    \label{fig:snr_adaptation}
\end{figure}

\section{Conclusion and Remarks}

In this work, we introduce a hybrid JSCC scheme for MIMO CSI compression that combines an autoencoder-based initial estimation with diffusion-based refinement. Our residual diffusion approach initializes reverse diffusion with a coarse autoencoder estimate, thereby tailoring it specifically for reconstruction. This two-stage framework can be exited after either stage based on channel conditions and performance requirements, balancing computational complexity against fidelity. Further, using \textit{$\chi$-prediction}, we enable a low-complexity 2-step diffusion inference with minimal performance loss, substantially reducing latency.
Our experiments across varying channel complexities show diffusion-based refinement delivers optimal value for complex channel distributions, while autoencoder solutions efficiently serve simpler channel distributions. These findings highlight the potential of diffusion models to enhance CSI reconstruction at base stations under challenging uplink conditions.
Future work includes developing multi-rate compression within a single model, investigating quantization effects, exploring weight-quantization techniques, and optimizing the U-Net backbone to improve computational efficiency while preserving performance. Further, including additional loss terms, such as the binary cross-entropy loss between transmitted and received data or the effective MIMO channel capacity, to directly optimize for the downstream task~\cite{li2023task} is also an interesting direction. 

\section*{Acknowledgment}
The authors thank Charlie Zhang for insightful discussions. This work was partly supported by Samsung Research America through the 6G$@$UT center within the Wireless Networking and Communications Group (WNCG) at the University of Texas at Austin, ARO Award W911NF2310062, ONR Award N000142412542, and NSF Award 2443857. 

\vspace{-.8em} 
 \medskip
 \small
 \bibliographystyle{IEEEtran}
 \bibliography{bibilography}

\clearpage
\normalsize



\end{document}